

\font\titlefont = cmr10 scaled\magstep 4
 2
\font\sectionfont = cmr10
\font\littlefont = cmr5 
\font\eightrm = cmr8

\def\ss{\scriptstyle}
\def\sss{\scriptscriptstyle}

\newcount\tcflag
\tcflag = 0  

\ifnum\tcflag = 0 \magnification = 1200 \fi  

\global\baselineskip = 1.2\baselineskip 
\global\parskip = 4pt plus 0.3pt 
\global\abovedisplayskip = 18pt plus3pt minus9pt
\global\belowdisplayskip = 18pt plus3pt minus9pt
\global\abovedisplayshortskip = 6pt plus3pt
\global\belowdisplayshortskip = 6pt plus3pt

\def\barsoff{\overfullrule=0pt}


\def\endignore{}
\def\ignore #1\endignore{} 

\newcount\dflag
\dflag = 0


\def\monthname{\ifcase\month 
\or January \or February \or March \or April \or May \or June%
\or July \or August \or September \or October \or November %
\or December 
\fi}

\newcount\dummy
\newcount\minute  
\newcount\hour
\newcount\localtime
\newcount\localday
\localtime = \time
\localday = \day

\def\advanceclock#1#2{ 
\dummy = #1
\multiply\dummy by 60
\advance\dummy by #2
\advance\localtime by \dummy
\ifnum\localtime > 1440 
\advance\localtime by -1440
\advance\localday by 1
\fi}

\def\settime{{\dummy = \localtime %
\divide\dummy by 60%
\hour = \dummy 
\minute = \localtime%
\multiply\dummy by 60%
\advance\minute by -\dummy 
\ifnum\minute < 10
\xdef\spacer{0} 
\else \xdef\spacer{}
\fi %
\ifnum\hour < 12
\xdef\ampm{a.m.} 
\else
\xdef\ampm{p.m.} 
\advance\hour by -12 %
\fi %
\ifnum\hour = 0 \hour = 12 \fi 
\xdef\timestring{\number\hour : \spacer \number\minute%
\thinspace \ampm}}}



\def\endtitle{}
\def\title#1\endtitle{\vskip.5in\titlefont
\global\baselineskip = 2\baselineskip 
#1\vskip.4in
\baselineskip = 0.5\baselineskip\rm}

\def\endauthors{}
\def\authors#1\endauthors{#1}

\def\endabstract{}
\def\abstract#1\endabstract{\vskip .3in%
\centerline{\sectionfont\bf Abstract}%
\vskip .1in
\noindent#1}

\def\nopageonenumber{\footline={\ifnum\pageno<2\hfil\else
\hss\tenrm\folio\hss\fi}}  

\newcount\nsection 
\newcount\nsubsection 

\def\section#1{\global\advance\nsection by 1
\nsubsection=0
\bigskip\noindent\centerline{\sectionfont \bf \number\nsection.\ #1}
\bigskip\rm\nobreak}

\def\subsection#1{\global\advance\nsubsection by 1
\bigskip\noindent\sectionfont \sl \number\nsection.\number\nsubsection)\
#1\bigskip\rm\nobreak}

\def\topic #1{{\medskip\noindent $\bullet$ \it #1:}}
\def\endtopic{\medskip}

\def\appendix#1#2{\bigskip\noindent%
\centerline{\sectionfont \bf Appendix #1.\ #2} 
\bigskip\rm\nobreak} 


\newcount\nref 
\global\nref = 1 

\def\therefs{}


\def\ref#1#2{\xdef #1{[\number\nref]} 
\ifnum\nref = 1\global\xdef\therefs{\item{[\number\nref]} #2\ } 
\else
\global\xdef\oldrefs{\therefs}
\global\xdef\therefs{\oldrefs\vskip.1in\item{[\number\nref]} #2\ }%
\fi%
\global\advance\nref by 1
}

\def\listrefs{\vfill\eject\section{References}\therefs}


\newcount\nfoot 
\global\nfoot = 1 

\def\foot#1#2{\xdef #1{(\number\nfoot)} 
\footnote{${}^{\number\nfoot}$}{\vbox{\baselineskip=10pt
\eightrm #2}}
\global\advance\nfoot by 1
}


\newcount\nfig 
\global\nfig = 1
\def\thefigs{} 

\def\figure#1#2{\xdef #1{(\number\nfig)}
\ifnum\nfig = 1\global\xdef\thefigs{\item{(\number\nfig)} #2\ }
\else
\global\xdef\oldfigs{\thefigs}
\global\xdef\thefigs{\oldfigs\vskip.1in\item{(\number\nfig)} #2\ }%
\fi%
\global\advance\nfig by 1 } 

\def\fig#1{\xdef #1{(\number\nfig)}
\global\advance\nfig by 1 } 


\newcount\ntab
\global\ntab = 1

\def\table#1{\xdef #1{\number\ntab}
\global\advance\ntab by 1 } 


\newcount\cflag
\newcount\nequation
\global\nequation = 1
\def\eqlabel{(1)}

\def\nexteqno{\ifnum\cflag = 0
\global\advance\nequation by 1
\fi
\global\cflag = 0
\xdef\eqlabel{(\number\nequation)}}

\def\lasteqno{\global\advance\nequation by -1
\xdef\eqlabel{(\number\nequation)}}

\def\label#1{\xdef #1{(\number\nequation)}
\ifnum\dflag = 1
{\escapechar = -1
\xdef\draftname{\littlefont\string#1}}
\fi}

\def\clabel#1#2{\xdef\eqlabel{(\number\nequation #2)}
\global\cflag = 1
\xdef #1{\eqlabel}
\ifnum\dflag = 1
{\escapechar = -1
\xdef\draftname{\string#1}}
\fi}

\def\cclabel#1#2{\xdef\eqlabel{#2)}
\global\cflag = 1
\xdef #1{\eqlabel}
\ifnum\dflag = 1
{\escapechar = -1
\xdef\draftname{\string#1}}
\fi}


\def\eeq{}

\def\eqnn #1\eeq{$$ #1 $$}

\def\eq #1\eeq{
\ifnum\dflag = 0
{\xdef\draftname{\ }}
\fi 
$$ #1
\eqno{\eqlabel \rlap{\ \draftname}} $$
\nexteqno}







\def\eqa #1\eeq{
\ifnum\dflag = 0
{\xdef\draftname{\ }}
\fi 
$$ \eqalignno{ #1 } $$
\global\cflag = 0}


\def\eg{{\it e.g.\/}}
\def\etc{{\it etc.\/}}
\def\etal{{\it et.al.\/}}


\def\anp#1#2#3{{\it Ann.\ Phys. (NY)} {\bf #1} (19#2) #3}

\def\npb#1#2#3{{\it Nucl.\ Phys.} {\bf B#1} (19#2) #3}

\def\pla#1#2#3{{\it Phys.\ Lett.} {\bf #1A} (19#2) #3}
\def\pra#1#2#3{{\it Phys.\ Rev.} {\bf A#1} (19#2) #3}
\def\prb#1#2#3{{\it Phys.\ Rev.} {\bf B#1} (19#2) #3}

\def\prd#1#2#3{{\it Phys.\ Rev.} {\bf D#1} (19#2) #3}
\def\pr#1#2#3{{\it Phys.\ Rev.} {\bf #1} (19#2) #3}
\def\prep#1#2#3{{\it Phys.\ Rep.} {\bf #1} (19#2) #3}
\def\prl#1#2#3{{\it Phys.\ Rev.\ Lett.} {\bf #1} (19#2) #3}
\def\prs#1#2#3{{\it Proc.\ Roy.\ Soc.} {\bf #1} (19#2) #3}
\def\rmp#1#2#3{{\it Rev.\ Mod.\ Phys.} {\bf #1} (19#2) #3}


\global\nulldelimiterspace = 0pt



\def\frac#1#2{{{#1} \over {#2}}\,}  
\def\hf{{1\over 2}}
\def\nth#1{{1\over #1}}


\def\Asl{\hbox{/\kern-.7500em\it A}} 
\def\Dsl{\hbox{/\kern-.6700em\it D}} 
\def\dsl{\hbox{/\kern-.5300em$\partial$}}
\def\pxpsl{\hbox{/\kern-.5600em$p$}}
\def\sslsh{\hbox{/\kern-.5300em$s$}}
\def\epssl{\hbox{/\kern-.5100em$\epsilon$}}
\def\delsl{\hbox{/\kern-.6300em$\nabla$}}
\def\lxpsl{\hbox{/\kern-.4300em$l$}}
\def\elxpsl{\hbox{/\kern-.4500em$\ell$}}
\def\kxpsl{\hbox{/\kern-.5100em$k$}}
\def\qxpsl{\hbox{/\kern-.5000em$q$}}
\def\sla#1{\raise.15ex\hbox{$/$}\kern-.57em #1}



\def\roughly#1{\mathrel{\raise.3ex
\hbox{$#1$\kern-.75em\lower1ex\hbox{$\sim$}}}}
\def\lsim{\roughly<}
\def\gsim{\roughly>}



\def\bfj{{\bf j}}

\def\bfn{{\bf n}}

\def\bfx{{\bf x}}



\def\Sce{{\cal E}}
\def\Scf{{\cal F}}

\def\Sch{{\cal H}}

\def\Scl{{\cal L}}

\def\Scn{{\cal N}}

\def\Scq{{\cal Q}}

\def\Scv{{\cal V}}


\def\ssb{{\sss B}}

\def\ssf{{\sss F}}

\def\ssn{{\sss N}}

\def\ssq{{\sss Q}}

\def\ssS{{\sss S}}
\def\sst{{\sss T}}

\def\ssv{{\sss V}}


\def\diag#1{{\rm diag}\left( #1 \right)}



\def\Avg#1{\left\langle #1 \right\rangle}



\def\hc{{\rm h.c.}}


\def\meV{{\rm \ meV}}
\def\eV{{\rm \ eV}}


\input epsf.tex

\nopageonenumber
\baselineskip = 18pt
\barsoff


\font\bigtitlefont    = cmr10 scaled\magstep 2
\def\BigTc{$\hbox{\titlefont T}_{\hbox{\bigtitlefont c}}$}

\def\HTc{{high-$T_c$}}
\def\Tc{{$T_c$}}
\def\TN{{$T_\ssn$}}

\def\Tx{{$T-x$}}

\def\five{{$SO(5)$}}
\def\four{{$SO(4)$}}
\def\three{{$SO(3)$}}
\def\two{{$SO(2)$}}
\def\twothree{{$SO(3) \times SO(2)$}}
\def\threetwo{\twothree}

\def\dt{{\partial_t}}

\def\nq{{n_\ssq}}
\def\ns{{n_\ssS}}
\def\Nq{{N_\ssq}}
\def\Ns{{N_\ssS}}
\def\eps{\epsilon}
\def\veps{\varepsilon}

\def\AF{{\sss AF}}
\def\SC{{\sss SC}}

\def\GB{{\sss GB}}
\def\pGB{{\sss pGB}}
\def\inv{{\rm inv}}
\def\sb{{\rm sb}}
\def\em{{\rm em}}
\def\spin{{\rm spin}}
\def\opt{{\rm opt}}
\def\dwave{$d_{x^2-y^2}$-wave}

\def\bk{\item{}}

\def\Nbar{\overline{N}}


\line{cond-mat/9705216 \hfil McGill-97/05 }

\title
\centerline{SO(5) Invariance and Effective Field}
\centerline{Theory for High \BigTc\ Superconductors}
\endtitle

\authors
\centerline{C.P. Burgess${}^a$ and C.A. L\"utken${}^b$}
\vskip .1in
\centerline{\it ${}^a$ Physics Department, McGill University}
\centerline{\it 3600 University St., Montr\'eal, Qu\'ebec, Canada, H3A 2T8.}
\vskip .05in
\centerline{\it ${}^b$ Physics Department, University of Oslo}
\centerline{\it P.O. Box 1048, Blindern, N-0316 Norway.}
\endauthors



\abstract
We set up the effective field theories which describe
the \five-invariant picture of the \HTc\ cuprates in various regimes.
We use these to get {\it quantitative} conclusions concerning the size
of \five-breaking effects.
We consider two applications in detail: ($i$) the thermodynamic
free energy, which describe the phase diagram and critical behaviour,
and ($ii$) the Lagrangian governing the interactions of the
pseudo-Goldstone bosons with each other and with the electron
quasiparticles deep within the ordered phases. We use these
effective theories to obtain predictions for  the critical behaviour
near the possible bicritical point and the pseudo-Goldstone
boson dispersion relations, as well as some preliminary
results concerning their contribution to response functions. We systematically
identify which predictions are independent of the microscopic details
of the underlying electron dynamics, and which depend on
more model-dependent assumptions.
\endabstract


\vfill\eject

\section{Introduction and Summary}

One of the remarkable features of \HTc\ cuprate
materials is the basic connection they display between
antiferromagnetism (AF) and superconductivity (SC). Indeed, any
convincing theory of these materials must explain
this fundamental property.

\ref\zhang{S.-C. Zhang, {\it SO(5) Quantum Nonlinear $\sigma$ Model
Theory of the High $T_c$ Superconductivity}, Stanford preprint,
cond-mat/9610140.}

S.-C.~Zhang \zhang\ has recently argued for a new perspective on
this AF-SC connection within the cuprates. He
identifies an approximate \five\ symmetry of the one-band
Hubbard model which
is believed to describe the dynamics of the pairing electrons within
the $Cu-O$ planes of these materials. This \five\ symmetry contains
as subgroups the \three\ symmetry of spin rotations (which is
spontaneously broken in the AF phase) and the electromagnetic
\two\ invariance (whose breaking defines the SC phase).
{}From Zhang's vantage point both ordered phases arise once \five\
is spontaneously broken, and the competition between
antiferromagnetism and superconductivity becomes a `vacuum
alignment' problem, in which the direction of the order parameter is
fixed by small effects which explicitly break the approximate
\five\ symmetry.

\ref\nscattexps{H.A. Mook et al., \prl{70}{93}{3490};\bk
J. Rossat-Mignod et al., Physica (Amsterdam) {\bf 235C}, 59 (1994);\bk
H.F. Fong et al., \prl{75}{95}{316}.}

\ref\pseudos{S. Weinberg, \prl{29}{72}{1698}.}

\ref\pGBHub{E. Demler and S.C. Zhang, \prl{75}{95}{4126};\bk
S. Meixner, W. Hanke, E. Demler and S.C. Zhang, Standford preprint,
cond-mat/9701217.}

Evidence for the validity of this picture comes from the observation
within the SC phase \nscattexps\ of a collective mode, centred near
momentum $(\pi/a,\pi/a)$, which couples to the spin-flip
channel in neutron-scattering experiments. Its energy gap
depends on doping in the same way as does \Tc\ itself,
taking the value $\Sce \simeq 41 \meV$ at optimal doping.
This state is understood in
Zhang's framework as a pseudo-Goldstone boson (pGB)
\pseudos, whose gap is kept small compared to the underlying
electron-electron interaction energy, $J \simeq 0.1 \eV$,
by the \five\ symmetry. Such a state is argued to be an approximate
eigenstate of the Hubbard model Hamiltonian in refs.~\pGBHub.

\ref\Wbg{S. Weinberg, Physica {\bf 96A} (1979) 327.}

As is true for {\it any} physical system \Wbg, the implications
of the \five\ picture at energies much lower than its
intrinsic scale, $J$, may be
efficiently encoded in terms of a low-energy effective theory. The
effective theory which does so will depend on the symmetries and
degrees of freedom which arise in the low-energy regime, and so can
differ depending on the regimes of temperature or doping which
are of interest.

Of particular interest are those low-energy predictions
which depend only on the symmetries and degrees of freedom
of the low-energy theory, and not, say, on the values of the
effective low-energy coupling constants. This is because
such predictions are robust, in the sense that they are
independent of the details of how these symmetries are realized
by the underlying electron dynamics.
This robustness obviously raises
the stakes of the comparison with experiment, since
disagreements cannot be attributed to small changes in microscopic
details.

\ref\ourletter{C.P. Burgess and C.A. L\"utken, preprint McGill-96/45
(cond-mat/9611070).}

Our goal in the present paper is to pursue the ideas of ref.~\zhang\
by systematically exploring the low-energy effective theories
which describe it in several regimes. A short summary of some of our
results has been given in ref.~\ourletter.  We pause here to list
some of our conclusions, before going into more detailed discussions
in subsequent sections.

\topic{(1) The Phase Diagram}
Our first application is to the system's phase diagram. In \S2, we
consider the system's free energy, and use it to
reiterate Zhang's description of the qualitative features of the
phase diagram in the temperature-doping (\Tx) plane. We then extend
this reasoning to argue that the {\it size} of the phases in this plane
are related to the size of the \five-breaking interactions, and
so to the gaps for the pseudo-Goldstone states.

More precisely, suppose the system Hamiltonian (or Lagrangian)
density has the form $\Sch = \Sch_\inv + \eps \, \Sch_\sb$, where
$\Sch_\inv$ is \five\ invariant, and $\eps \ll 1$ quantifies
the magnitude of the explicit \five-breaking interactions at
zero doping. We argue below that the gap for the
pseudo-Goldstone state in the AF phase is $\Sce_\AF = O(\eps^{1/2} J)$,
while that in the SC phase is $\Sce_\SC = O(\eps^{1/3} J)$. We
learn the size of $\eps$ from the observed 41 meV gap of the
SC phase: $\eps \sim 0.01$. Both the N\'eel temperature
at zero doping (\TN), and the SC critical temperature at optimal
doping (\Tc), are predicted to to be of order $\eps^{1/2}  J$,
leading to the (correct) expectation that these should be
of order $100 K$.

Similar statements hold for the doping required at zero temperature
to destroy the AF order and enter the SC regime. This transition is
predicted for dopings of order $x_c \simeq \eps^{1/2} \simeq 10\%$.
In addition, a potential `mixed' phase (MX), which is both AF and SC in
nature, can arise between the purely SC and AF phases, again over
a range of dopings which is of order $\eps^{1/2} \simeq 10\%$. The
superconducting phase itself is expected to extend out to dopings
for which the \five\ invariance is no longer a good approximation.

\ref\surveytexts{For summaries of typical phase diagrams see, \eg,
J.W. Lynn, {\it High Temperature Superconductivity}, Springer-Verlag,
1990; \bk
G. Burns, {\it High Temperature Superconductivity, An Introduction},
Academic Press, 1992.}

Although the $O(1)$ prefactors to these estimates are model
dependent, and so will vary from material to material, the general
shape of the resulting phase diagram should be shared by all
\HTc\ systems, and seems to agree reasonably well with the typical
experimental phase diagram \surveytexts.

\topic{(2) Critical Behaviour}
The \HTc\ systems differ from traditional superconductors in that
they display critical behaviour within a few degrees of \Tc. This
critical behaviour is the topic of \S2.3.  The
\five-invariant scheme predicts the usual scaling behaviour for
the transitions from the SC or AF regimes into the disordered phase.
It makes different predictions for the vicinity of a potential
bicritical (or tetracritical) point which is expected should
the SC and AF phases coexist for some temperatures and dopings.
Zhang has argued that one is attracted here to an \five-invariant fixed
point, based on the $d=2+\veps$ expansion for $d$-dimensional
nonlinear sigma-models. We point out here an alternative possibility
where the stable fixed point weakly breaks \five, based on the
$d=4-\veps$ expansion, and compute the critical exponents in
this picture.

\topic{(3) Pseudo-Goldstone Boson Spectrum}
Next, \S3 applies \five\ to the underlying Hamiltonian for energies,
temperatures and dopings corresponding to the ordered phases,
and obtains the properties of the pseudo-Goldstone modes.
As was emphasized elsewhere \ourletter, since
there are more features to these gap spectra than there are
parameters in the general effective theory, it is possible to predict
low-energy relations amongst the various features of the dispersion
relations for these states.

Model-independent, weak-coupling conclusions do not appear
to be possible for the disordered phase, although inferences
regarding this phase may be made by making more specific use
of additional assumptions concerning the microscopic dynamics.

Our pGB Lagrangian differs in detail from
that proposed by Zhang, since it involves a few more terms.
Although they completely agree for qualitative purposes, we argue
the inclusion of the most general interactions is required to
permit the robust inference of the quantitative size of
symmetry-breaking effects.

\topic{(4) Response Functions}
Finally, \S4 uses the effective theory to make some preliminary
points concerning the pseudo-Goldstone boson contributions to
electromagnetic and spin response functions within
the AF and SC phases. In particular, the origin within the
effective theory, of the pseudo-Goldstone
pole in the SC-phase spin response is identified.
\endtopic

\section{Implications for Thermodynamics}

In terms of the underlying electrons, the ordered AF and SC
phases of the \HTc\ systems are distinguished by nonzero values
for the order parameters  $\Avg{\psi_{p} (i\sigma^2)
\psi_{-p}}$ and $\Avg{\psi^\dagger_{p+Q} \sigma^a \psi_{p}}$,
where $p$ is an electron momentum,
$Q = ({\pi \over a}, {\pi \over a})$ and $\sigma^a$ denotes
the Pauli matrices.
Long-wavelength variations of these quantities can therefore
be described by the fields:
\label\eoparams
\eq
\nq(k) \propto \psi_{p+k} (i \sigma_2) \psi_{-p}
\qquad \hbox{and} \qquad
n^a_\ssS(k) \propto \psi^\dagger_{p+Q+k}
\sigma^a \psi_{p} ,
\eeq
where $k$ is much smaller than either $p$ or $Q$.

In Zhang's framework, in addition to the \threetwo\ symmetry
of spin and electromagnetic gauge transformations there are
approximate symmetries which rotate $\nq$ and $\ns$,
into one another. To represent this symmetry it is convenient to
group $\nq$ and $\ns$ into a real, five-dimensional quantity
\label\fivevec
\eq
\bfn = \pmatrix{\nq \cr \ns \cr},
\eeq
on which the
extended symmetry acts by matrix multiplication on the left
by an orthogonal five-by-five matrix: $\bfn \to O \; \bfn$,
where $O$ is an arbitrary five-by-five orthogonal matrix.
This identifies the electromagnetic \two\ and spin \three\
subgroups as the block-diagonal combinations:
\label\embedding
\eq
O = \pmatrix{ SO(2) & 0 \cr 0 & SO(3) \cr}.
\eeq
We denote by $T_\alpha$ the hermitian,
antisymmetric five-by-five
matrices which generate \five, with the special cases of
the generators of \two\ and \three\ represented by:
\label\charges
\eq
Q = q \pmatrix{\sigma^2 &\cr & 0 \cr}, \qquad \hbox{and} \qquad
\vec T = \pmatrix{0 & \cr & \vec t \cr} .
\eeq
Here $q=2$ is the electric charge of the order parameter in units
of the electron charge and $\vec t$ are a basis of three-by-three
generators of \three.

The resulting \five\ symmetry is simultaneously
subject to two kinds of breaking. On one hand it is broken
explicitly (but weakly) to the exact \twothree\ subgroup by
the electron Hamiltonian, and on the other hand it is
spontaneously broken to \four. What follows in this section
spells out our assumptions concerning how this symmetry
manifests itself in thermodynamic functions.

\subsection{Thermodynamic Potentials}

\ref\zhangmac{D. Arovas, A.J. Berlinsky, C. Kallen and S.C. Zhang,
preprint cond-mat/9704048.}

Given this picture we may write the free-energy density,
$f$, as the sum of an \five-invariant term, $f_\inv$, and a
small \five-breaking term, $f_\sb$.  For a slowly varying
order parameter we may use a derivative expansion of $f$,
leading to the following most general expressions for time-independent
$\nq$ and $\ns$:\foot\newfeatures{This expression differs
from that of refs.~\zhang, \zhangmac\ in three ways.
Besides working in the isotropic limit, these authors:
(1) impose the constraint $\ss \ns\cdot\ns+\nq\cdot\nq=1$,
(2) keep only three
of the lowest terms in an expansion of the $\ss w$'s
and $\ss u$'s in powers of fields,
and (3) keep only quadratic terms in the potential $\ss v$. (Their
quartic interactions arise from the chemical-potential dependence
of the terms involving two derivatives of $\ss \nq$.)
We comment further on these
differences as they arise at subsequent points in the text.}
\label\invform
\eq
\eqalign{
\beta f_\inv &= v_\inv  +  w^{(+)}_\parallel \Bigl(
\nabla_a {\nq} \cdot \nabla_a {\nq} +
\nabla_a {\ns} \cdot \nabla_a {\ns} \Bigr) +
u^{(+)}_\parallel \Bigl( {\nq} \cdot \nabla_a {\nq} +
{\ns} \cdot \nabla_a {\ns} \Bigr)^2 \cr
& \qquad + w^{(+)}_\perp \Bigl(
\nabla_c {\nq} \cdot \nabla_c {\nq} +
\nabla_c {\ns} \cdot \nabla_c {\ns} \Bigr) +
u^{(+)}_\perp \Bigl( {\nq} \cdot \nabla_c {\nq} +
{\ns} \cdot \nabla_c {\ns} \Bigr)^2  + \cdots , \cr
\beta f_\sb &= v_\sb   +  w^{(-)}_\parallel \Bigl(
\nabla_a {\nq} \cdot \nabla_a {\nq} -
\nabla_a {\ns} \cdot \nabla_a {\ns} \Bigr) +
u^{(-)}_\parallel \Bigl( {\nq} \cdot \nabla_a {\nq} -
{\ns} \cdot \nabla_a {\ns} \Bigr)^2 \cr
& \qquad+  w^{(-)}_\perp \Bigl(
\nabla_c {\nq} \cdot \nabla_c {\nq} -
\nabla_c {\ns} \cdot \nabla_c {\ns} \Bigr) +
u^{(-)}_\perp \Bigl( {\nq} \cdot \nabla_c {\nq} -
{\ns} \cdot \nabla_c {\ns} \Bigr)^2 \cr
&\qquad\qquad\qquad + u^{(0)}_\parallel  \; {\nq} \cdot \nabla_a {\nq}
\; {\ns} \cdot \nabla_a {\ns} + u^{(0)}_\perp  \; {\nq} \cdot
\nabla_c {\nq} \; {\ns} \cdot \nabla_c {\ns}
+ \cdots  . \cr}
\eeq
Here $\beta = 1/kT$, and the ellipses represent terms
involving more derivatives than two.
The coefficient functions, $v_\inv, v_\sb, w^{(\pm)}_\perp,
w^{(\pm)}_\parallel,  u^{(\pm,0)}_\perp$ and $u^{(\pm,0)}_\parallel$
are potentially arbitrary functions of the \threetwo\ invariants
$\nq \cdot \nq$ and $\ns \cdot \ns$. They depend as well on the
two thermodynamic variables: temperature and
doping (more about this dependence below). Keeping in mind the
anisotropy of the cuprate systems, separate coefficients
are included for derivatives, $a=x,y$, parallel to the
copper oxide planes, and those, $c=z$, perpendicular to this plane.
For simplicity we assume rotational symmetry within the planes.

In the limit of \five\ invariance, $v_\sb = w^{(-)}_\perp =
w^{(-)}_\parallel = u^{(-)}_\perp = u^{(-)}_\parallel = u^{(0)}_\perp
= u^{(0)}_\parallel = 0$,
and each of the remaining functions $v_\inv, w^{(+)}_\perp,
w^{(+)}_\parallel, u^{(+)}_\perp$ and $u^{(+)}_\parallel$ depend
only on the combination $\bfn \cdot\bfn = \nq \cdot \nq +
\ns \cdot \ns$.

The predictive power of
the approximate symmetry comes from computing observables
perturbatively in the small symmetry-breaking interactions.
Doing so requires a quantitative characterization of the size of
the symmetry-breaking terms in $ f$. We imagine, therefore,
the microscopic electronic Hamiltonian
to have the generic form:
\label\genericmicro
\eq
\Sch = \Sch_\inv + \eps \; \Sch_\sb - \mu \; \Scq,
\eeq
where $\Sch_\inv$ preserves \five, while $\Sch_\sb$ and $\mu\;
\Scq$ both break it. $\mu$ here is the chemical potential for
the electric charge, $\Scq$. We take
$\mu=0$ to correspond to half filling.

Notice that \five\ symmetry is explicitly broken in two different
ways in eq.~\genericmicro, and so there are two different
parameters, $\eps$ and $\mu$, which govern the
symmetry-breaking terms in the free energy, $f$:

\topic{Zero Doping}
Things are simplest at half filling, for which $\mu=0$ and so only
one \five-breaking parameter exists: $\eps$.
Based on our later discussion of the spectrum of the gap in the
pseudoGoldstone boson spectrum in the SC phase, we take in what
follows $\eps\simeq  0.01$. It is this small parameter which
ultimately determines the size of the different phases in the system's
phase diagram.

\topic{Nonzero Doping}
The second source of explicit \five\ breaking in eq.~\genericmicro\
is the chemical potential, $\mu$, which is introduced to describe
the doping of \HTc\ systems away from half filling. This breaks \five\
because $\mu$ couples only to electric charge, which is just
one of the ten \five\ generators. For thermodynamic
applications it is more convenient to trace the dependence of the
free energy on the number density of charge carriers,
$n$, rather than $\mu$,
and so we define the doping, $x$, by $n = x/\Scv$, with $\Scv$
the volume of a unit cell.  In typical \HTc\ systems at zero
temperature, $x \lsim 10\%$ is characteristic of the AF phase,
while $0.1 \lsim x \lsim 0.4$ is representative of the SC phase
\surveytexts.
\endtopic

To explore the implications of these two sources of symmetry breaking
we consider first the classical phase transitions which are implied by
the choice of an \five\ invariant order parameter.

\subsection{Classical Phase Transitions}

\ref\gbrgcrit{For a review of the critical behaviour of the \HTc\
systems, see Q. Li in {\it Physical Properties of High-Temperature
Superconductors}, Volume $V$, edited by D.M. Ginsberg, World
Scientific, 1996.}

For a classical phase transition the free energy density,
$ f(x,T,\nq,\ns)$,
is assumed to admit a Taylor expansion in powers of $\nq$ and $\ns$
near $\nq=\ns=0$. This assumption of analyticity
at zero field is known to fail in the vicinity of
a critical point, where the long-distance fluctuations can cause $ f$
to acquire a singular dependence on its arguments. For
temperatures closer to \Tc\ than a few degrees, the
resulting critical behaviour is poorly described by a classical
discussion. Unlike traditional superconductors, the correlation
length in \HTc\ systems is sufficiently small to permit critical
behaviour to be observed,  although typically only within a
few degrees of \Tc\  \gbrgcrit.

For these reasons we expect a classical analysis to be adequate for
the purposes of identifying the overall size occupied by the various
phases within the phase diagram in the \Tx\ plane. We investigate the
implications of \five\ for the
critical behaviour in \S2.3, below.

\ref\correcn{We thank G.R. Zemba, as well as M.A. Martin-Delgado and
J.R. Pelaez for correspondence concerning this point.}

We start by rederiving Zhang's treatment of classical phase transitions.
Our addition to his argument is the quantifying of the size
of the symmetry-breaking effects in this analysis. Consider therefore
expanding the potential $v$ to quartic order
in the order parameters, $\nq$ and
$\ns$:\foot\newnotation{This potential corrects a minor
error \correcn\ in the potential in the original version of
ref.~\ourletter, although none
of the results of this reference are altered by this change.}
\label\quarticorder
\eq
v = v_{00} + \hf \Bigl( v_{20} \; \nq\cdot\nq + v_{02} \;
\ns\cdot\ns \Bigr) +
\nth{4} \Bigl[  v_{40} \; (\nq\cdot\nq)^2 + v_{22} \;
\nq\cdot\nq \ns\cdot\ns +
v_{04} \; (\ns\cdot\ns)^2 \Bigr] + \cdots,
\eeq
where the coefficients, $v_{jk}$, are regarded as functions of
the thermodynamic variables, $x$ and $T$, as well as the
small symmetry-breaking parameter $\eps$. Particle-hole
symmetry\foot\xtomnx{Alternatively, the same conclusion
is also drawn in later sections by considering the free energy
to be a function
of $\ss \mu$ rather than $\ss x$, and using the antisymmetry of
$\ss Q$. See also \ourletter. } further implies $v_{jk}$
must also satisfy $v_{jk}(-x) = v_{jk}(x)$, if $T$ and $\eps$
are held fixed.

Next divide this expression into its \five-invariant and -breaking
parts, $v = v_\inv + v_\sb$, where:
\label\invandsbterms
\eq
\eqalign{
v_\inv &= v_{00} + {\rho \over 2} \; \Bigl( \nq\cdot\nq + \ns\cdot\ns \Bigr) +
{\lambda \over 4} \; \Bigl( \nq\cdot\nq + \ns\cdot\ns \Bigr)^2 + \cdots, \cr
v_\sb &= {g \over 2} \; \Bigl( \nq\cdot\nq - \ns\cdot\ns \Bigr)
+ {k \over 4} \; \Bigl( \nq\cdot\nq - \ns\cdot\ns \Bigr)^2
+ {h \over 4} \; \Bigl[  (\nq\cdot\nq)^2 - (\ns\cdot\ns)^2 \Bigr] +
\cdots . \cr}
\eeq
The parameters in these expressions are related as follows:
$v_{02} = \rho - g$, $v_{20} = \rho + g$, $v_{40} = \lambda + k + h$,
$v_{04} = \lambda + k - h$ and $v_{22} = \lambda - k$.

\fig\phasediagone

\midinsert
$$\vbox{\tabskip=0pt \offinterlineskip
\halign to \hsize{\strut#& #\tabskip 1em plus 2em minus .5em&
\hfil#\hfil &#&\hfil$#$\hfil &#& \hfil$#$\hfil &#\tabskip=0pt\cr
\noalign{\hrule}\noalign{\smallskip}\noalign{\hrule}\noalign{\medskip}
&& Phase && \hbox{Minimum} && \hbox{Conditions}&\cr
\noalign{\medskip}\noalign{\hrule}\noalign{\medskip}
&&  Normal (N) && |\nq| = |\ns| = 0 && v_{20}>0,
v_{02}>0 &\cr
&& Superconductor (SC) && |\nq|\ne 0, |\ns|=0 &&
v_{20}<0, v_{02}>0 &\cr
&& Antiferromagnet (AF) && |\nq|=0, |\ns|\ne 0 &&
v_{20}>0, v_{02}<0 &\cr
&& Mixed (MX) && |\ns|\ne 0, |\nq| \ne 0 &&
v_{20}<0, v_{02}<0, k >0  &\cr
\noalign{\medskip}\noalign{\hrule}\noalign{\smallskip}
\noalign{\hrule}
}}$$
\centerline{\bf Table I}
\medskip
\noindent \vbox{\baselineskip=10pt \eightrm
The possible minima of the free energy density,
$\ss v$, each distinguished by the sign of $\ss v_{02}$
and $\ss v_{20}$. Two possibilities arise if both of
these parameters are negative, depending on the sign taken by
the quartic coupling, $\ss k$. $\ss k>0$, defines a minimum for
which both $\ss |\ns|$ and $\ss |\nq|$ are nonzero. If $\ss k<0$,
then there is both AF and SC type minima, separated by an
energy barrier.  The corresponding phase diagram in the $\ss \rho-g$
plane is given for both choices for $\ss k$ in Fig.~\phasediagone.}
\endinsert

Minimization of this potential gives four kinds of phases depending
on whether $\ns$ and/or $\nq$ vanish at the minimum. These four
alternatives are controlled, in first approximation, by the signs of
the coefficients of the two quadratic terms, $v_{20}$ and
$v_{02}$, as outlined in Table I. The case where both
$v_{20}$ and $v_{02}$ are negative, further subdivides
into two alternatives which are distinguished by the sign of the
quartic coupling, $k$. If $k>0$ then both $|\nq|$ and $|\ns|$
are nonzero at the minimum, while if $k<0$ then two types of
minima coexist, which differ according to whether it is $|\ns|$
or $|\nq|$ which is nonzero. When $k<0$ these two minima are
separated by an energy barrier.

\midinsert
\centerline{\epsfxsize=9.5cm\epsfbox[0 430 450 750]{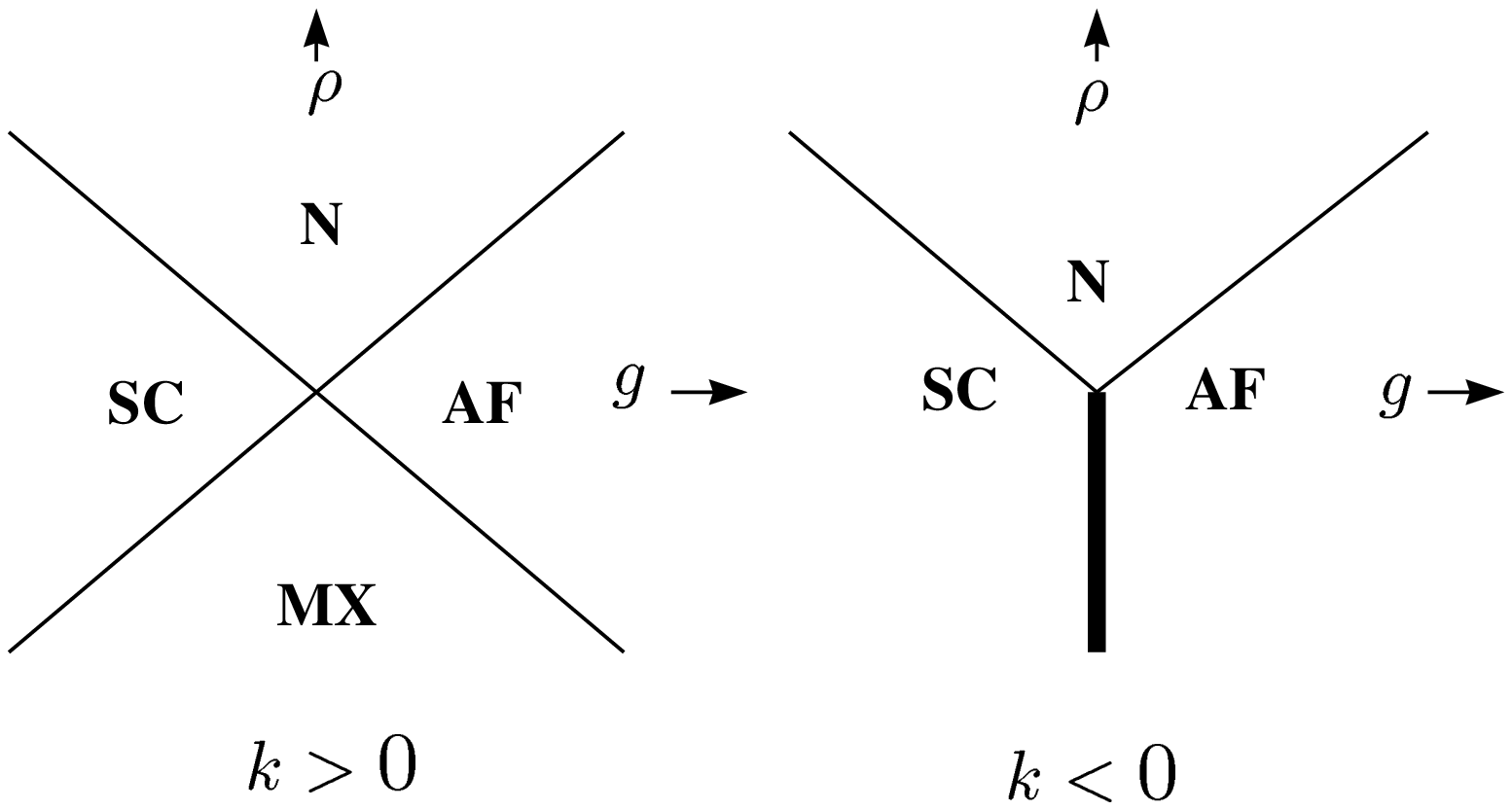}}

\bigskip
\centerline{\bf Figure 1}
\medskip\noindent
\vbox{\baselineskip=10pt \eightrm The phase diagram for the free energy
function
described in the text in the $\rho-g$ plane, where $\rho$ is the
\five-symmetric, and $g$ is the \five-breaking, quadratic couplings
in the free energy. The thin lines represent second-order transition
lines, while the fat line is a first-order phase boundary.
The two possibilities portrayed are distinguished
by the sign of the quartic coupling, $k$, in the free energy.}
\bigskip
\endinsert

The corresponding phase diagram, in the $\rho-g$ plane, is drawn
in Fig.~\phasediagone. When $k>0$ there are four phases, and
all phase boundaries define second-order
transitions. The four phase boundaries intersect at a tetracritical
point, which is defined by $\rho=g=0$. When $k<0$ there are only
three phases, with the antiferromagnetic (AF) and superconducting
(SC) regions meeting at a first-order transition. This phase boundary
intersects the second-order boundaries with the disordered phase (N)
at a bicritical point, again at $\rho=g=0$.
The picture which emerges is the semiclassical phase diagram
described in ref.~\zhang.

More information becomes available, however,
once the dependence of the
transition regions on the small symmetry-breaking quantity $\eps
\simeq 0.01$ is included. To see this, imagine examining $v$ in
the vicinity of half filling, where $x \ll 1$. Keeping in mind that
all \five-breaking interactions must vanish if both $x$ and $\eps$ do,
and that $v_{jk}$ must be even functions of $x$, we
write in this limit:
\label\invcexp
\eq
g = \eps \, g_0 + g_2 \, x^2 + g_4 \, x^4 + \cdots, \qquad
h = \eps \, h_0 + h_2 \, x^2 + h_4 \, x^4 + \cdots, \qquad
k = \eps \, k_0 + k_2 \, x^2 + k_4 \, x^4 + \cdots ,
\eeq
where the coefficients of this expansion are functions of $T$
which can be unsuppressed by additional powers of $\eps$ or $x$.

Since the \five-invariant couplings need not vanish with either
$\eps$ or $x$, for them we instead write:
\label\invcexp
\eq
\lambda = \lambda_0 + \lambda_2 \, x^2  + \lambda_4 \, x^4 + \cdots .
\eeq

The \five-invariant quadratic term must be handled more carefully,
however. This is because real \HTc\ systems are antiferromagnets
at zero temperature and doping, and so we have the important
additional information $v_{20}(T=x=0) > 0$ and $v_{02}(T=x=0) < 0$.
Equivalently this implies: $g(T=x=0) > 0$ and
$-g(T=x=0) < \rho(T=x=0) < g(T=x=0)$,
which is only consistent with $g(T=x=0) = O(\eps)$ if $\rho(T=x=0)$
is also $O(\eps)$. We take, therefore, at zero temperature:
\label\rhoform
\eq
\rho = \eps \, \rho_0 + \rho_2 \, x^2 + \rho_4 \, x^4 + \cdots .
\eeq

\topic{Transition Dopings}

These expansions determine the domains of doping over which the
various phases are possible, so long as $x$ is small. For temperatures below
the N\'eel temperature -- and so for which $-g_0 < \rho_0 < g_0$ -- define
$x_\AF$ as the doping above
which the AF order is lost. $x_\AF$ is then determined by the
requirement that $v_{02} = \rho - g = 0$, for which eqs.~\invcexp\
and \rhoform\ imply:
\label\xafexpr
\eq
x^2_\AF = - \eps \; \left({ \rho_0 - g_0 \over \rho_2 - g_2}\right) +
O(\eps^2).
\eeq
So long as both numerator and denominator are both order unity, and
$\rho_2 > g_2$, we see the prediction is:
$x_\AF = O({\eps^{1/2}}) \simeq 10\%$, which agrees well with what
is observed for real \HTc\ systems.

Similarly, a superconducting phase arises for dopings greater than
$x_\SC$, defined as the value of $x$ for which $v_{20}=\rho + g = 0$.
Again eqs.~\invcexp\ and \rhoform\ give:
\label\xscexpr
\eq
x^2_\SC = - \eps \; \left( { \rho_0 + g_0 \over \rho_2 + g_2} \right)
+ O(\eps^2) ,
\eeq
and so $x_\SC$ is also predicted to be $O(\eps^{1/2}) \simeq 10\%$,
so long as numerator and denominator are $O(1)$ and
$\rho_2 + g_2 < 0$.

If we instead work to $O(x^4)$ in $v_{20}$, then a second root $x_\SC'$
can develop beyond which superconductivity is again lost. As is easy
to see, this second root arises for $x \simeq O(1)$,
and so lies outside of the domain of the small-$x$ expansion, and for
$x$ potentially large enough to invalidate the approximate \five\
symmetry. Notice that this implies that the optimal doping,
$x_\opt$ --- defined as the doping for which \Tc\ is largest --- is
also $O(1)$, and so is unsuppressed by powers of $\eps$.

Finally, if $x_\AF > x_\SC$, and if $k>0$, then dopings satisfying
$x_\SC < x < x_\AF$ give both a superconductor and an
antiferromagnet (as in the MX phase) at zero temperature. This
phase can extend over a range of dopings which is at most as large
as $x_\AF - x_\SC = O({\eps^{1/2}})$. Otherwise,
if $k<0$, the AF to SC transition is first order.

\topic{Transition Temperatures}
Similar arguments may be applied to estimate the size
of the critical temperatures for the various ordered phases if the
$T$ dependence of the coefficients in $v$ is known. In mean-field
theory the large-temperature limit of the temperature dependence
of the free energy may be argued on dimensional grounds. For large $T$
in $d$ dimensions, the quadratic coefficients, $\rho$ and $g$, are
proportional to $T^2$, with a prefactor that can be $O(1)$ for $\rho$,
but which is suppressed by $\eps$ or $x^2$ for $g$.

For sufficiently large $T$ it follows that eventually $\rho > |g|$,
implying a transition to the disordered phase (N). The transition
temperature as a function of doping may be estimated by asking
when the $T^2$ contribution is the same size as is the
zero-temperature values of for $\rho$ and $g$.
Since the zero-temperature limits of both $\rho$ and $g$ are
suppressed by powers of $\eps$ and/or $x^2$, the transition temperature, $T_c$,
is predicted to be smaller than might be
expected based on the underlying electronic scales, $J \simeq
0.1 \eV \simeq 1000 \, K$. At zero
doping only $\eps$ controls the size of symmetry breaking,
so the N\'eel temperature, $T_\ssn$, is expected to be
\label\Tcest
\eq
T_\ssn = O(\eps^{1/2} J) \simeq 100 \, K .
\eeq
Once again this is the right order of
magnitude for the cuprates. Of course, these estimates can
be further reduced by other effects. For example, very
anisotropic systems are effectively two dimensional,
in which case $w_\perp^{(\pm)}, u^{(\pm,0)}_\perp \to 0$,
implying $T_c \to 0$.

\vskip-1in
\centerline{\epsfxsize=7.5cm\epsfbox[45 130 550 750]{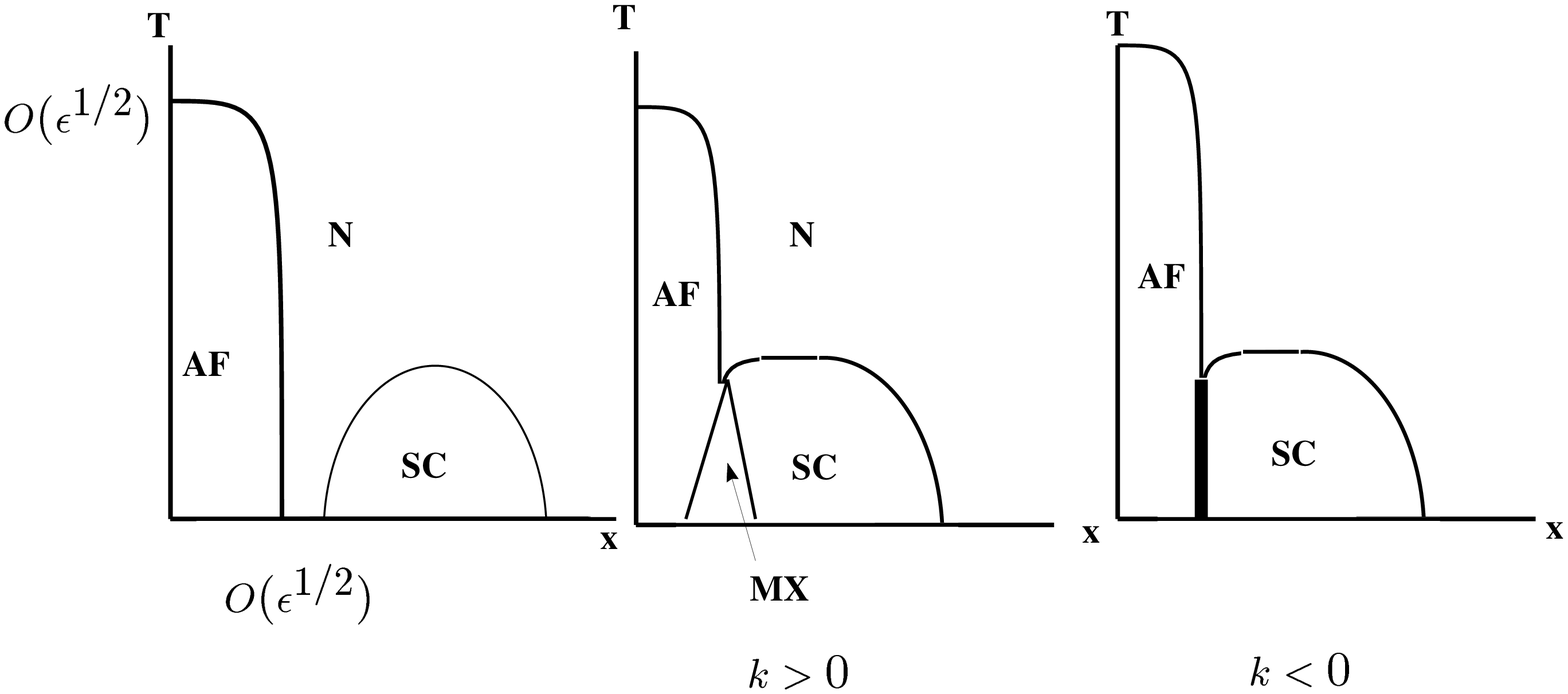}}

\bigskip
\centerline{\bf Figure 2}
\medskip\noindent
\vbox{\baselineskip=10pt \eightrm
The phase diagram for the free energy function
described in the text in the temperature--doping plane. As for
Fig.~\phasediagone, the thin lines represent second-order transition
lines, while the fat line is a first-order phase boundary.
The two possibilities for which the AF and SC phases coexist
are distinguished by the sign of the quartic coupling, $k$, in the
free energy.}
\bigskip

\fig\phasediagtwo

We are thus led to the phase diagram of Fig.~\phasediagtwo\ in the
temperature--doping plane, which reproduces well the generic
features of the phase diagram for real \HTc\ systems.
\endtopic

\vfill\eject
\subsection{Critical Behaviour}

Since the symmetry of the order
parameter has important implications for the critical exponents,
and since some critical behaviour is experimentally accessible
for the cuprates, we pause to explore critical phenomena more carefully here.

\ref\twodfp{M.E. Fisher and D.R. Nelson, \prl{32}{74}{1350};\bk
R.A. Pelcovits and D.R. Nelson, \pla{57}{76}{23}; \prb{16}{77}{2191};\bk
D.S. Fisher, \prb{39}{89}{11783};\bk
D.H. Friedan, \anp{163}{85}{318}.}

As is well known, in $d<4$ dimensions the Gaussian fixed point
is unstable in the infrared, and so the critical exponents are
controlled by some other infrared fixed point of the
renormalization group.  In the absence of more reliable methods,
one is reduced to expanding about either the upper ($d=4$) or
lower ($d=2$) critical dimensions. Our approach here is to
expand in powers of $d-4$, and we do so for both the critical
lines separating the disordered (N) phase from the others, as
well as for the bi- or tetra-critical point should these lines meet.
We imagine in what follows that all five of the fields in
$\ns$ and $\nq$ are free
to fluctuate. Our treatment of the tetracritical
point therefore differs from Zhang's, since he draws his conclusions
for this point using the nonlinear sigma model, for which $\ns\cdot\ns
+ \nq\cdot\nq =1$, based on an expansion about $d=2$ \twodfp.

\ref\scxyfp{M.E. Fisher, M.N. Barber and D. Jasnow,
\pra{8}{73}{1111};\bk
B.I. Halperin, T.C. Lubensky and S.-K. Ma, \prl{32}{74}{292};\bk
D.S. Fisher, M.P.A. Fisher and D.A. Huse, \prb{43}{91}{130}.}

Along the critical line the \five\ picture implies the usual
$O(N)$-invariant infrared fixed points, where $N=3$ for
the N\'eel transition, and $N=2$ (XY universality \scxyfp)
for the superconducting transition\foot\xyrider{Due to the relevance
of the electromagnetic coupling, the SC transition crosses over from
XY universality to that of a charged fluid, although undetectably close
to $\ss T_c$ \scxyfp.}. Both are covered by the
usual result to leading order in
the expansion in $\veps = 4-d$:
\label\onfp
\eq
{v_{*} \over 8 \pi^2} = {\veps \over N+8}
+ O(\veps^2), \qquad \nth\nu = 2 - {N+2 \over N+8} \;
\veps + O(\veps^2) ,
\eeq
where $v_*$ denotes the fixed-point value of
$v_{40}$ for the N/SC transition (or of $v_{04}$
for the N/AF transition).\foot\fieldrescale{Very near the critical point
(and so in this section only) we ignore the system's
anisotropy and rescale all fields to canonically normalize
the derivative terms in the free energy.}
$\nu$ is the standard critical exponent for the correlation
length, which diverges like $\xi \sim t^{-\nu}$ as $t
\equiv T/T_c -1 \to 0$. There is indeed good experimental
evidence \gbrgcrit\ that $\nu$ for the N/SC transition,
is given as predicted for XY universality.

\ref\compfp{See, \eg, P.M. Chaikin and T.C. Lubensky,
{\it Principles of Condensed Matter Physics}, Cambridge
University Press, 1995.}

The fixed point behaviour in the bicritical (or tetracritical) region differs
qualitatively from that found along the two critical lines. This is
because the $O(N)$-invariant fixed point is stable only for
$N<4-O(\veps)$, and so does not apply for $N=5$. The fixed
point appropriate for $N=5$ may be found numerically, and
the results obtained agree where they can be compared
with previous treatments \compfp,
as well as with the approximate analytic expressions given
below.

In order to identify the fixed points it is useful to consider a
model for which the field $\ns$ spans an $\Ns$-dimensional
space, and $\nq$ spans a space of $\Nq$ dimensions. The
free energy is assumed to be $O(\Ns)\times O(\Nq)$
invariant. In this case useful approximate expressions for
the fixed points may be derived in an expansion in powers
of $n \equiv \Ns - \Nq$ over $\Nbar=\Ns+\Nq$.
In the case of present interest $\Ns=3$ and $\Nq=2$, and so
the parameter $n/\Nbar=1/5$ is reasonably small.

The stable fixed points which emerge at leading order in
$\veps=4-d$ then are: ($i$) the $O(\Nbar)$-invariant
fixed point, for $\Nbar <4$; ($ii$) the `decoupled' fixed point (for
which $ (v_{22})_*=0$), for $(\Ns+2)(\Nq+2)>36$; and ($iii$) a third
fixed point for intermediate values of $\Ns$ and $\Nq$. It is
this third fixed point which is appropriate for $\Ns=3$ and
$\Nq=2$, and it gives:
\label\fixedpointpos
\eq
{\lambda_*\over 4 \pi^2} \approx 8 \; \lambda_0 ,
\qquad {h_*\over 4 \pi^2} \approx {n \, (\Nbar-4) \over \Nbar - 2}
\; \lambda_0 ,
\qquad {k_*\over 4 \pi^2} \approx 2 (\Nbar-4) \; \lambda_0 ,
\eeq
where $\lambda_0 \equiv \veps / (\Nbar^2 + 32)$. For $\Nbar=5$,
$n=1$ and $d=3$ these expressions agree well with the values
we obtained numerically to linear order in $\veps$:
$\lambda_*/(4\pi^2) \approx 0.141$, $h_*/(4 \pi^2)
\approx 0.00573$, and $k_*/(4 \pi^2) \approx 0.0340$.
Notice the hierarchy $h_* \ll k_* \ll
\lambda_*$ which is satisfied at this fixed point, and is consistent
with approximate \five\ invariance, albeit with some
symmetry-breaking
couplings which are larger than $O(1\%)$ of their \five-invariant
counterparts.

Since the picture of how the critical lines merge near the
bicritical point differs somewhat from that of ref.~\zhang,
we next present this in more detail.
The running of the two quadratic couplings, $\rho$ and $g$,
in the far infrared is given near the fixed point by:
\label\rgbfns
\eq
\eqalign{
\left(\mu {\partial \rho \over \partial \mu}\right)_*
&= -2 \, \rho + {1\over 16 \pi^2} \;
\Bigl\{ \Bigl[ 2(\Nbar+2)\lambda_* +4k_* - n h_* \Bigr] \, \rho
+ \Bigl[ (\Nbar+4)h_* - 2n\lambda_* \Bigr] \, g \Bigr\} ,\cr
&= -2 \, \rho + {\lambda_0\over 4(\Nbar-2)} \; \Bigl\{ \Bigl[24 \Nbar
(\Nbar-2) - n^2(\Nbar-4) \Bigr]\, \rho + \Bigl[(\Nbar-8)^2-48
\Bigr] \, n\, g \Bigr\} ,\cr
\left(\mu {\partial g \over \partial \mu}\right)_*
&= -2 \, g + {1\over 16 \pi^2} \;
\Bigl\{ \Bigl[(\Nbar+4)h_* - 2n k_* \Bigr]\, \rho
+\Bigl[ 4\lambda_* + 2(\Nbar+2)k_* - nh_*\Bigr] \, g \Bigr\} ,\cr
&= -2 \, g + {\lambda_0\over 4(\Nbar-2)} \; \Bigl\{ -3n(\Nbar-4)^2
\, \rho + \Bigl[4\Nbar(\Nbar-2)^2 - n^2(\Nbar-4) \Bigr] \, g \Bigr\} .
\cr}
\eeq
The principal directions of the flow, and the corresponding scaling
exponents, near the fixed point are given to good approximation by:
\label\diagops
\eq
\rho_+ \approx \rho - \left({3 n (\Nbar-4)^2 \over 4 \Nbar (\Nbar-2)
(8-\Nbar)}\right) \; g \qquad
\rho_- \approx g - \left( { n [(\Nbar-8)^2
- 48] \over 4 \Nbar(\Nbar-2)(8-\Nbar)} \right) \; \rho
\eeq
and:
\label\diagexps
\eq
{1\over \nu_+} \approx 2 - {6 \Nbar \over \Nbar^2 + 32} \; \veps,
\qquad\qquad
{1\over \nu_-} \approx 2 - {\Nbar (\Nbar-2) \over
\Nbar^2 + 32} \; \veps .
\eeq
For $\Nbar=5$ and $n=1$ these become $\rho_+ \simeq
\rho - 0.017 \; g$ and $\rho_- \simeq g + 0.22 \; \rho$, while
$\nu_+ \simeq 0.68$ and $\nu_- \simeq 0.58$.

The largest of these exponents defines the temperature exponent,
$\nu = \nu_+ \simeq 0.68$. The scaling form for the free energy density,
$f = t^{-d\nu} \, \Scf\left({\rho_- \over t^\phi}
\right)$, defines the crossover exponent, $\phi$, which is
given by $\phi = \nu_+/\nu_-$. We have:
\label\fpnuval
\eq
\phi \approx {\Nbar^2 + 2\Nbar + 64 \over 2(\Nbar^2 -
3\Nbar + 32)} \simeq 1.2 .
\eeq
Since $\phi > 1$, the
two critical lines approach the bicritical point tangent to one
another, and to the line $\rho_-=0$ --- or, equivalently, to the
line $g = \left( { n [(\Nbar-8)^2 - 48] \over 4 \Nbar(\Nbar-2)(8-\Nbar)}
\right) \; \rho \simeq 0.22 \, \rho$.

It would obviously be of great interest to examine experimentally
the critical behaviour at the bicritical point, since this would
more decisively distinguish the symmetries of the order parameter
at this point. Unfortunately, to our knowledge, there is currently
no experimental information concerning this critical behaviour,
likely due to the disorder which is introduced in this region
by the doping process.

\section{The Nonlinear Sigma Model: Pseudo-Goldstone Modes}

\ref\chiral{The formalism for describing the dynamics of Goldstone
and pseudoGoldstone modes was developed within the context of
the strong interactions by:\bk
S. Weinberg, \prl{18}{67}{188};
\pr{166}{68}{1568}; {\it op. cit.};\bk
S. Coleman, J. Wess and B. Zumino, \pr{177}{69}{2239};
E.C. Callan, S. Coleman, J. Wess and B. Zumino, \pr{177}{69}{2247};\bk
J. Gasser and H. Leutwyler, Ann. Phys. {\bf 158} (1984) 142;
Nucl. Phys. {\bf B250} (1985) 465.}

\ref\xptnotes{Two relatively recent reviews of the Goldstone boson
formalism, including applications to condensed matter physics,
are given by:\bk
H. Leutwyler, {\it Nonrelativistic Effective Lagrangians},
Bern preprint BUTP-93/25;\bk
C.P. Burgess, {\it An
Introduction to Effective Lagrangians and their Applications},
lecture notes for the Swiss
Troisi\`eme Cycle, Lausanne, June 1995.}

More microscopically, another regime for which \five-invariance
provides definite quantitative predictions is deep within the ordered
phases. Here the low-energy dynamics is governed by the Goldstone,
and pseudo-Goldstone modes, together with their interactions with
the other low-energy degrees of freedom. Much can be said about
these low-energy properties because low-energy Goldstone-boson
dynamics is largely determined purely by the pattern of
spontaneous symmetry breaking \Wbg, \chiral, \xptnotes.

The present section is devoted to developing the quantitative
description of this regime. Section \S3.1 first describes the most
general low-energy effective Lagrangian for this phase, which is
then used in \S3.2 to determine the dispersion relations for the
pseudo-Goldstone bosons themselves. Their interactions are used
in \S4 to draw preliminary conclusions about their contributions to
response functions in the AF and SC phases, which also involves
their interactions with the low-energy electron-like
quasiparticles.

\subsection{The Effective Lagrangian}

Imagine, then, that $T$ and $x$ are chosen to lie within one of
the ordered phases, and integrating out all degrees of freedom
which involve energies larger than some scale, $\Lambda$. We
wish to write down the effective Lagrangian governing the
degrees of freedom which remain at still lower energies.
Although this effective Lagrangian cannot yet be derived from
the microscopic physics, we are guaranteed to capture its physics
so long as we use the {\it most general possible} Lagrangian involving
the low-energy modes and respecting all of the symmetries
\Wbg.

When a global symmetry, $G$, is spontaneously broken to a
subgroup, $H$, the self-couplings and spectrum of the resulting
(pseudo-) Goldstone states is described at low energies
by the nonlinear sigma model for the quotient space
$G/H$ \Wbg, \chiral, \xptnotes. When $G=SO(5)$ and
$H=SO(4)$ this
implies the lowest terms in the derivative expansion of the
Lagrangian for this system are completely determined by two
constants. Of course, more possibilities arise once explicit
\five-breaking interactions are introduced.

The most general such Lagrangian involving two or fewer
derivatives is again built from the fields $\ns$ and $\nq$, but
with the important difference, relative to \S2.1, that these now satisfy the
constraint $\ns\cdot \ns + \nq \cdot \nq \equiv 1$, since we
are interested in only the Goldstone and pseudo-Goldstone modes.
In the absence of doping,
the result takes a form which is similar to eq.~\invform, $\Scl =
\Scl_\inv + \Scl_\sb$, with:\foot\notnpt{Our notation here follows
that of ref.~\ourletter, generalized to the anisotropic case. Other
couplings, such as $\ss \eps^{mnl} \partial_m \vec{n}_\ssS
\cdot (\partial_n \vec{n}_\ssS \times \partial_l \vec{n}_\ssS)$,
are also possible in specific dimensions (in this case $\ss d=2$).}
\label\newinvform
\eq
\eqalign{
\Scl_\inv &= {f_t^2 \over 2} \Bigl(\partial_t {\nq} \cdot
\partial_t {\nq} + \partial_t {\ns} \cdot \partial_t {\ns} \Bigr) -
 {f_\parallel^2 \over 2} \Bigl(\nabla_a {\nq} \cdot
\nabla_a {\nq} + \nabla_a {\ns} \cdot \nabla_a {\ns} \Bigr) \cr
&\qquad \qquad
+  {f_\perp^2 \over 2} \Bigl(\nabla_c {\nq} \cdot
\nabla_c {\nq} + \nabla_c {\ns} \cdot \nabla_c {\ns} \Bigr)
+ \cdots \cr
\Scl_\sb &= -V + f^2_t \Bigl[ A \, \partial_t {\nq} \cdot
\partial_t {\nq} + B \,\partial_t {\ns} \cdot \partial_t {\ns}
+ C\,  (\nq \cdot \partial_t \nq)^2 \Bigr] \cr
&\qquad\qquad - f_\parallel^2 \Bigl[ D_\parallel \nabla_a {\nq}
\cdot \nabla_a {\nq} + E_\parallel \nabla_a {\ns} \cdot
\nabla_a {\ns} + F_\parallel (\nq \cdot \nabla_a \nq)^2 \Bigr] \cr
&\qquad\qquad - f_\perp^2 \Bigl[ D_\perp \nabla_c {\nq}
\cdot \nabla_c {\nq} + E_\perp \nabla_c {\ns} \cdot
\nabla_c {\ns} + F_\perp (\nq \cdot \nabla_c \nq)^2 \Bigr]
+ \cdots ,\cr}
\eeq
where $f_t$, $f_\parallel$ and $f_\perp$ are constants, while
$V, A_{\parallel,\perp}, B_{\parallel,\perp}, C_{\parallel,\perp},
D_{\parallel,\perp}, E_{\parallel,\perp}$ and $F_{\parallel,\perp}$
are potentially arbitrary functions of the \threetwo\ invariants
$\nq \cdot \nq$ and $\ns \cdot \ns$. They also can depend on the
temperature, $T$, since this can appear in
$\Scl$ through the process of integrating out the high-energy modes.

Fewer terms appear in eq.~\newinvform\ than in eq.~\invform\
because of the constraint $\ns\cdot \ns +
\break \nq \cdot \nq = 1$ which
is enforced in \newinvform, but not in \invform. Should the `length'
of the \five-breaking order parameter also describe a propagating
mode in the low-energy system, as might be appropriate near
the critical point, then this constraint may be relaxed. Since this
mode is {\it not} guaranteed to be in the low-energy theory
far away from the phase boundaries --- unlike the
pseudo-Goldstone states --- we do not include it further
in this section.

\ref\finn{A. H\ae rdig and F. Ravndal, {\it Eur. J. Phys.} {\bf 14}
(1993) 171.}

Far from the critical region the correlation length perpendicular
to the planes becomes smaller than half the interplane spacing,
and we take the system to be approximately two-dimensional,
corresponding to the limit\foot\belayer{A more detailed
modelling of the dimensional crossover
within the context of Bose-Einstein condensation
is given in ref.~\finn.} $f_\perp  = 0$.
We imagine working in this limit in what follows, and so from here on
drop the redundant subscript `$\parallel$' from the coefficient functions
$D$, $E$ and $F$.

\ref\bosemu{Kothari and Singh, \prs{A178}{41}{135};\bk
P.T. Landsberg and J. Dunning-Davies, \pr{138}{65}{A1049};\bk
J.I. Kapusta, \prd{24}{81}{426};\bk
H.E. Haber and H.A. Weldon, \prl{46}{81}{1497}; \prd{25}{81}{502}.}

\ref\bosedispersion{J. Bernstein and S. Dodelson, \prl{66}{91}{683};\bk
K.M. Benson, J. Bernstein and S. Dodelson, \prd{44}{91}{2480}.}

Couplings to long-wavelength electromagnetic fields is
incorporated in the effective theory through the usual
substitution $\partial_\alpha \nq \to (\partial_\alpha - ie
A_\alpha Q)\nq$. (A special case of this coupling is the
dependence on the chemical potential, $\mu$, which enters
$\Scl$ through the replacement $\partial_t \nq \to
(\partial_t - ie\mu Q)\nq$ \bosemu, \bosedispersion.)
For distances shorter than the electromagnetic
screening length, $a = eq f_t$, this electromagnetic coupling
includes the Coulomb interactions of the pseudo-Goldstone
bosons. It is important to realize, however, that the strong
{\it short-ranged} Coulomb interactions need not be included
in this way, since these are integrated out to arrive
at $\Scl$ in the first place. Although
strong microscopic interactions such as these would complicate the derivation
of the $\Scl$ from first principles, they play no role
when using $\Scl$ at low energies.

An important consequence of these observations now follows.
When using the Lagrangian of eq.~\newinvform, the key observation
is that all of the interactions are {\it guaranteed} to be weak at low
energies, justifying a perturbative treatment. This is because all
interaction terms are suppressed by either a derivative or a small
\five-breaking parameter or both. In particular, the pseudo-Goldstone
spectrum may be obtained from $\Scl$ in mean field theory
by expanding in fluctuations about minima of the potential, $V$.

For later purposes a useful parameterization (which identically
solves the constraint $\nq\cdot\nq + \ns\cdot \ns = 1$) is given
by polar coordinates on the four-sphere:
\label\explangles
\eq
 \nq = \cos\theta  \pmatrix{\cos\phi \cr \sin\phi \cr},
\qquad\qquad
\ns = \sin\theta \pmatrix{ \sin\alpha \cos\beta \cr
\sin\alpha \sin\beta \cr \cos\alpha\cr} ,
\eeq
(although care is required to properly handle those points
where these coordinates are singular).
In terms of these variables, and including a chemical potential,
the Lagrangian becomes:
\label\kintermangles
\eq
\eqalign{
\Scl &= - V+  {f^2_t \over 2} \Bigl[ \Bigl(1 + 2A \sin^2\theta +
2 B \cos^2\theta + 2C\sin^2\theta\cos^2\theta \Bigr)
(\partial_t \theta)^2  \cr
&\qquad\qquad + (1 + 2A ) \cos^2\theta
(\partial_t \phi + eq\mu )^2
+ (1+2B)\sin^2\theta \Bigl( (\partial_t \alpha)^2
+ \sin^2\alpha (\partial_t \beta)^2 \Bigr) \Bigr] \cr
& \qquad \qquad - {f^2_\parallel \over 2} \Bigl[ \Bigl(1 + 2D
\sin^2\theta + 2 E \cos^2\theta + 2F\sin^2\theta\cos^2\theta
\Bigr)(\nabla_a \theta)^2 \cr
&\qquad\qquad + (1 + 2D ) \cos^2\theta (\nabla_a \phi)^2
+ (1+2E) \sin^2\theta \Bigl( (\nabla_a \alpha)^2
+ \sin^2\alpha (\nabla_a \beta)^2 \Bigr) \Bigr] + \cdots,\cr }
\eeq
where all coefficient functions are to be regarded
as functions of $\cos^2\theta$.

To this point we have not yet used much information concerning
the nature or size of the explicit symmetry breaking. This we now do
by making an assumption as to how the symmetry-breaking
terms transform under \five. Since there are two types of symmetry
breaking, a choice must be made for each.

\topic{Doping}
Doping has been incorporated into the effective theory through the
chemical potential, $\mu$. We return to the connection between
$\mu$ and $x$ in \S3.3 below.

\topic{Intrinsic Breaking}
Symmetry breaking also occurs at half filling, with strength $\eps$.
The resulting symmetry-breaking pattern is $SO(5) \to SO(3)\times
SO(2)$. We assume this to be done with the simplest possible `order
parameter', $M$, which transforms in the adjoint representation
of \five. That is, choose $M = \eps \, \diag{3,3,-2,-2,-2}$.
\endtopic

The Lagrangian is then the most general
function of the fields $\bfn = {\nq \choose \ns}$, $\mu Q$ and
$M$, subject to the following \five\ transformation
property
\label\transrule
\eq
\Scl(O \bfn, O \mu Q O^\sst, O M O^\sst) = \Scl(\bfn,\mu Q,M) ,
\eeq
where $O$ is an \five\ transformation.

The utility of identifying $Q$ and $M$ may be seen when
$\Scl$ is expanded in powers of the small quantities $\eps$ and
$\mu$. Since $M$ and $Q$ always appear premultiplied by these
small numbers, this expansion restricts the kinds of symmetry breaking which
can arise order by order, which in turn constrains the possible
$\theta$-dependence of the coefficient functions in $\Scl$.

For example, a term in the scalar potential involving $2n$ powers of
$\bfn$ must have the following form:
\label\potexpn
\eq
V_{(n)} = \sum_{(k_1,l_1) \ne (0,0)} \cdots \sum_{(k_n,l_n)\ne (0,0)}
C_{k_1l_1,\dots,k_n,l_n}\Bigl[ \bfn^\sst (\eps M)^{k_1}
(\mu Q)^{2l_1} \bfn \Bigr] \cdots
\Bigl[ \bfn^\sst (\eps M)^{k_n} (\mu Q)^{2l_n} \bfn \Bigr].
\eeq
Only even powers of $Q$ enter here due to its antisymmetry, and
the term $k_i = l_i = 0$ is excluded from the sums
due to the constraint $\bfn^\sst \bfn = 1$. Expanding
$\Scl$ to low order in the \five-breaking parameters
$\eps$ and $\mu$ necessarily also implies keeping only the lowest
powers of $\nq \cdot \nq = \cos^2\theta$ in $V$.

Similar conclusions may be obtained for the other
coefficient functions in the Lagrangian of eq.~\newinvform.
Working to $O(\eps^2,\eps\mu^2,\mu^4)$ in $V$, and to
$O(\eps,\mu^2)$ in the two-derivative terms
then gives:
\label\leadingpotterms
\eq
V = V_0 + V_2 \,\cos^2\theta + \hf \, V_4 \,\cos^4\theta,
\eeq
and
\label\leadingkinterms
\eq
\eqalign{
&A = A_0 + A_2 \cos^2\theta, \qquad B = B_0 + A_2 \cos^2\theta,
\qquad C = C_0, \cr
&D= D_0 + D_2 \cos^2\theta, \qquad
E= E_0 + D_2 \cos^2\theta, \qquad
F= F_0 , \cr}
\eeq
for the coefficient functions in eq.~\newinvform. Notice that the
terms proportional to $\cos^2\theta$ in $A$ and $B$ are identical,
as are the corresponding terms in $D$ and $E$.
Expanding in powers of $\eps$ and $\mu$, the constants
in eqs.~\leadingpotterms\ and \leadingkinterms\
start off linear in $\eps$ and $\mu^2$: $A_i = A_i^{10}
\eps + A_i^{01}\mu^2 + \cdots$ \etc.
The only exceptions to this statement are: $B_0, E_0 \propto \eps$
(no $\mu^2$ term), $C_0, F_0 \propto \mu^2$ (no $\eps$ term),
and $V_4 = V_4^{20} \eps^2 + V_4^{11} \eps \mu^2 + V_4^{02} \mu^4$.
Furthermore, since the $\mu^2 \; \nq\cdot \nq$ term in $V$
arises from
substituting $\dt \to \dt -i e\mu Q$ in the kinetic term for $\nq$,
we have: $V_2^{01} = - \, \hf \, f_t^2 e^2 q^2$ to leading order.
Higher powers of $\mu$ originate from terms in $\Scl$ which involve
more than two derivatives.

In this way we arrive at an effective lagrangian very similar to
that of  ref.~\zhang\ (in the isotropic limit). The main difference
here is the power counting of the symmetry breaking terms.
Ref.~\zhang\ keeps terms quadratic in $\nq$ and has three
susceptibility parameters controlling the time-derivative terms,
and three stiffness constants governing the spatial derivatives.
Although a quadratic scalar potential captures the leading order
in $\eps$, it does not appear that a three-parameter derivative
term corresponds to any fixed order in $\eps$ or $\mu^2$.

\subsection{Pseudo-Goldstone Dispersion Relations}

We now turn to the calculation of the pseudo-Goldstone
boson dispersion relations. The scalar potential of
eq.~ \kintermangles\ has three types of extrema:
$$\eqalign{
&(1) \qquad \theta_0 = 0 \quad\hbox{or}\quad \pi ; \cr
&(2) \qquad \theta_0 = {\pi \over 2} \quad\hbox{or}
\quad {3\pi \over 2} ; \cr
&(3) \qquad \theta_0 \quad \hbox{where $c = \cos\theta_0$
satisfies $V'(c^2) = 0$} . \cr}$$
This leads to the four classical ordered phases found in
\S 2.2: ($i$) SC phase:
extremum (1) is a minimum, and (2) is a maximum; ($ii$)
AF phase: (2) is a minimum, and (1) is a maximum; ($iii$) MX phase:
both (1) and (2) are maxima, and (3) is a minimum; or ($iv$)
metastable phase: both (1) and (2) are minima, and (3) is a
maximum. This analysis becomes identical to that of \S 2.2
if $V$ is assumed to be quartic in $\cos\theta$, as was done
in ref.~\ourletter.

\topic{Superconducting Phase}
An expansion about the superconducting mimimum, $\theta_0=0$,
gives the dispersion relations in this phase
for the four bosons. The result
is a spin triplet of pseudo-Goldstone modes for which
\label\simpdispform
\eq
E(k) = \Bigl[ c^2 \, k^2 + \Sce^2 \Bigr]^\hf,
\eeq
with the phase speed, $c^2_\pGB(SC)$, and gap, $\Sce^2_\pGB(SC)
\equiv \Sce^2_\SC$, given to lowest order in \five-breaking
parameters by:
\label\SCpseudos
\eq
\eqalign{
c^2_\pGB(SC) &= {f^2_\parallel \over f^2_t} \;
\Bigl[ 1 + 2 \Bigl(E(1)-B(1)\Bigr) \Bigr]  \cr
&= {f^2_\parallel \over f^2_t} \; \Bigl[ 1 + 2 \Bigl(E_0-B_0\Bigr)
+ 2 \Bigl(D_2-A_2\Bigr) \Bigr], \cr
\Sce^2_\SC &= { -2 V'(1)  \over f^2_t} \cr
&= { -2 (V_2 + V_4)  \over f^2_t} .\cr}
\eeq
In both of these results the first equation uses the general
effective theory, eq.~\kintermangles, while the second equality
incorporates the additional information of eqs.~\leadingpotterms\
and \leadingkinterms.

The remaining field, $\phi$, describes a {\it bona fide}
gapless Goldstone mode. Its dispersion relation, $E(k)$ is a
more complicated function of $c^2 k^2$ and $eq\mu$ whose
form \bosedispersion, is not required here. Its phase
velocity, $c^2 \equiv c_\GB^2(SC) $, is given by
\label\SCGB
\eq
\eqalign{
c^2_\GB(SC) &= {f^2_\parallel \over f^2_t} \;
\Bigl[ 1 + 2 \Bigl( D(1)-A(1) \Bigr) \Bigr] \cr
&= {f^2_\parallel \over f^2_t} \;
\Bigl[ 1 + 2 \Bigl(D_0-A_0\Bigr) + 2\Bigl(D_2 - A_2 \Bigr)\Bigr]  . \cr}
\eeq
Recall in these expressions that $V(\cos^2\theta)$ includes any
$\mu$-dependent contributions coming from the kinetic terms,
or their higher-derivative counterparts, and a prime denotes
differentiation with respect to $\cos^2\theta$.

\topic{Antiferromagnetic Phase}
Expanding about the AF minimum gives the usual two
magnons satisfying dispersion relation \simpdispform, with:
\label\AFmagnons
\eq
\eqalign{
c^2_\GB(AF) &= {f^2_\parallel \over f^2_t} \;
\Bigl[ 1 + 2 \Bigl(E(0)-B(0)\Bigr) \Bigr] \cr
&= {f^2_\parallel \over f^2_t} \;
\Bigl[ 1 + 2 \Bigl(E_0-B_0\Bigr) \Bigr] , \cr
\Sce^2_\GB(AF) &= 0 . \cr}
\eeq
The remaining two states form a pair of electrically-charged
pseudo-Goldstone bosons satisfying:
\label\compdispform
\eq
E_\pm(k) = \Bigl[ c^2\, k^2 + \Sce^2 \Bigr]^\hf \pm e q\mu,
\eeq
with:
\label\AFpseudos
\eq
\eqalign{
c^2_\pGB(AF) &= {f^2_\parallel \over f^2_t} \;
\Bigl[ 1 + 2 \Bigl(D(0)-A(0)\Bigr)  \Bigr] \cr
&= {f^2_\parallel \over f^2_t} \;
\Bigl[ 1 + 2 \Bigl(D_0-A_0\Bigr)  \Bigr],  \cr
\Sce^2_\AF = \Sce^2_\pGB(AF) &= { 2 V'(0) \over f^2_t}  \cr
&= { 2 V_2 \over f^2_t} . \cr }
\eeq
\endtopic

These expressions imply the simple formulae of ref.~\ourletter:
$\Sce^2_\AF = m^2 - \kappa \mu^2$, and $\Sce^2_\SC = -m^2
+\kappa\mu^2 - \xi \mu^4$, where $m^2 \equiv 2 V^{10}_2
\eps/f_t^2 + O(\eps^2)$, $\kappa \equiv -2 V_2^{01}/f_t^2 +
O(\eps) = e^2q^2 + O(\eps)$ and $\xi \equiv 2 V_4^{02}/f_t^2 +
O(\eps)$.

Within the AF phase the
pseudo-Goldstone boson gap is seen to fall linearly
with $\mu^2$: $\Sce^2_\AF \approx \Sce^2_\AF(0)
[\mu^2_\AF - \mu^2]$,
where $\mu_\AF$ represents the doping for which one leaves
the AF regime. Similarly $\Sce^2_\SC$ varies quadratically
with $\mu^2$. By eliminating parameters of the Lagrangian
in favour of properties of the gap as a function of $\mu$ we
find the relations of ref.~\ourletter:
\label\vepspreds
\eq
\eqalign{
\varepsilon^2_\AF(\mu) &= {\varepsilon_\AF^2(0) \over \mu^2_\AF}
\, \Bigl[ \mu^2_\AF - \mu^2 \Bigr] ,\cr
\varepsilon^2_\SC(\mu) &= {\varepsilon^2_\SC(\opt) \over
\mu^4_\opt } \, (\mu^2
- \mu^2_{\SC-}) (2\mu^2_\opt - \mu^2) ,\cr
{\varepsilon_\AF^2(0) \over \mu^2_\AF} &= 2 \;
{\varepsilon^2_\SC(\opt) \over
\mu^2_\opt } ,\cr
\mu^2_\AF &= \mu^2_{\SC-} + O(\epsilon^2) , \cr}
\eeq
where $\mu_\opt$ here denotes the chemical potential
corresponding to the maximum gap, $\Sce_\SC$. We expect this
to occur at optimal doping, $\mu_\opt = \mu(x_\opt)$.

Similarly, the phase velocities for all modes in both SC and
AF phases are equal to one another, and to $f^2_t/f^2_\parallel$,
in the strict \five-invariant limit. (The parameters $f_t$ and
$f_\parallel$ are related to the compressibility and magnetic
penetration depth in the next section.) The $O(\eps)$ corrections to this limit
also satisfy some model-independent relations, which follow by
eliminating parameters from the above expressions:
\label\speedrels
\eq
c^2_\phi(SC) - c^2_\phi(AF) = c^2_\alpha(SC) - c^2_\alpha(AF) =
O(\eps).
\eeq

\subsection{The Connection Between $x$ and $\mu$}

The previous expressions giving the dependence of physical
quantities in terms of the chemical potential, $\mu$, would be more
useful if expressed in terms of the physically-measured
quantity, the doping, $x$. This relation is determined implicitly
in the present section.

The dependence $\mu(x)$ is found by adjusting $\mu$ to
ensure that the net electric charge equals $x$ charge carriers
per unit cell.  This must be done differently in the AF and SC phases.

\ref\dwaveev{For a review of the evidence for \dwave\
pairing in the \HTc\
systems, see J.F. Annett, N. Goldenfeld and A.J. Leggett,
 in {\it Physical Properties of High-Temperature
Superconductors}, Volume $V$, edited by D.M. Ginsberg, World
Scientific, 1996; \bk
D.J. Scalapino, \prep{250}{95}{329}.}

In both phases the total charge is carried by both the charged
pseudo-Goldstone states, {\it and} the ordinary electron-like
quasiparticles responsible for conduction. The existence of
these electron-like quasiparticles at low energies in the SC
phase is demanded by the evidence in favour of a
\dwave\ gap in the cuprate superconductors \dwaveev.
These experiments
argue for the existence of ungapped states due to the nodes of
the \dwave\ gap function, which cannot be provided by the
four pseudo-Goldstone modes. In the AF phase these degrees
of freedom correspond to ordinary unpaired electrons.

\topic{AF Phase}
In the AF phase the charge is carried by a mixture of electrons and
charged pseudo-Goldstone states. The condition of equilibrium
between these two types of charge carriers implies their electric
chemical potentials must be equal, and so the defining condition
for $\mu(x)$ becomes:
\label\nvsxAF
\eq
\Avg{\rho_\em} \equiv {ex \over \Scv} = \int_0^\infty d\omega \;
\Bigl\{
\Scn_\ssf(\omega) \Bigl[ n_\ssf(\mu) - n_\ssf(-\mu) \Bigr]
+ \Scn_\ssb(\omega) \Bigl[ n_\ssb(\mu) - n_\ssb(-\mu) \Bigr] \Bigr\},
\eeq
where $\Scn_{\ssf}$ and $\Scn_{\ssb}$ respectively
denote the density of states for fermions and bosons. For weakly
interacting bosons in $d$ spatial dimensions, whose dispersion
relation is $E^2 = k^2 c^2 + \Sce^2$, the density of states
is given explicitly by $\Scn_\ssb = \left({\Omega_d/(2\pi)^{d} }\right)
\; k^{d-2}E/c^2$, where $\Omega_d$ is the solid angle swept out by a
vector in $d$ spatial dimensions (so $\Omega_2 = 2\pi$ and
$\Omega_3 = 4\pi$). The Bose-Einstein distributions are given by
$n_\ssf(\mu) = \Bigl[ e^{(\omega - e\mu)/kT} + 1 \Bigr]^{-1}$ and
$n_\ssb(\mu) = \Bigl[ e^{(\omega - qe\mu)/kT} - 1 \Bigr]^{-1}$.

For small $\mu$ it is the fermions which dominantly
contribute to $\rho$, and the dependence $x(\mu)$ which results
is (for $d=2$ space dimensions) linear: $x \propto \mu/J$.
This linear dependence changes once $\mu$ becomes of order
the scalar gap at zero doping, $\Sce_\AF(0) = O(\eps^{1/2} J)$,
since at this point the scalar charge density varies strongly
with $\mu$, signalling the transition to the condensed (SC) phase.

\S2.2 showed the doping for which one exits (at low temperature) from
the AF phase to be, in order of magnitude, $x=x_\AF = O(\eps^{1/2})$.
We see that this corresponds to a chemical potential whose size is
$\mu \equiv \mu_\AF = O(\eps^{1/2} J)$.
(An identical conclusion regarding the size of $\mu_\AF$ may
be drawn from the condition for Bose-Einstein condensation:
$\mu \simeq \Sce_\AF(0)$.)

We see that since the chemical potential depends linearly
on doping within the AF regime, eqs.~\vepspreds\ implies
a linear dependence of $\Sce^2_\AF(x)$ on $x^2$:
\label\AFgapvsx
\eq
\Sce_\AF^2(x) = \Sce^2_\AF(0)(x_\AF^2 - x^2) .
\eeq
Corrections to this linearity in $x^2$ arise as the SC phase is
approached since $\mu$ no longer varies linearly with $x$
near the point where the bosons condense.

\topic{SC Phase}
Deep within the SC phase we assume the charge density to
be dominated by the $\nq$ condensate. Writing the potential
as $V(n) \approx \hf \, (m^2 - \nu^2) n^2+ {1 \over 24} \, g^2 n^4$
(where $\nu \equiv eq \mu f_t $,
$m^2  \equiv 2V_2(\mu=0) \approx 2 V^{10}_2 \eps$,
and $g^2 = 12 V_4(\mu=0) \approx 12 V^{20}_4 \eps^2$), we see
$\Avg{\rho_\em}$ is given by
\label\nvsxSC
\eq
\Avg{\rho_\em} \equiv {eqx \over \Scv} = - \; \left. {\partial V
\over \partial \mu}
\right|_{n = n_0} = {6 eqf_t \nu \over g^2} \, \Bigl( \nu^2 -  m^2 \Bigr) ,
\eeq
where $n_0 = 6 (\nu^2 - m^2)/g^2$ minimizes $V$.
This leads to the asymptotic expressions
$\nu \approx  \left( {g^2 x / 6 \Scv f_t} \right)^{1/3} \propto x^{1/3}$
if $g^2  \Avg{\rho_\em}/(6 f_t m^3) \gg 1$, and $\nu - m
\approx gx/ (12 \Scv f_t m^2 )\propto x$ if $g^2  \Avg{\rho_\em}/
(6 f_t m^3) \ll 1$.

In the case of interest we have (for $d=2$ space dimensions)
$f_t =  O(1)$, $m^2 = O(\eps J^2)$,
and $g^2 = O(\eps^2 J)$, and so $g^2  \Avg{\rho_\em}/(6 f_t m^3)
\simeq x  \eps^{1/2}/(6 \Scv J^2)$. Taking $J \simeq 0.1\, \eV$ and
$\Scv \simeq (10 \AA)^2$ we find $6 \Scv J^2 \simeq 10^{-5}$, leading to
$g^2  \Avg{\rho_\em}/(6 f_t m^3) \simeq 10^5 x  \eps^{1/2}
\simeq 10^4 x$. Since this is much greater than unity when
$x \gsim x_\SC = O(\eps^{1/2})$,
it follows that $\mu \propto x^{1/3}$ within the SC phase.

We may now determine the size of $\mu$ at optimal doping. We
found in \S2.2 that optimal doping occurs for $x = x_\opt =
O(1)$, and so
$x_\opt/x_\AF = O(\eps^{- \, \hf})$. But since $\mu
\propto x^{1/3}$ within the SC phase, we see
$\mu_\opt \sim \mu_\AF (x_\opt/x_\AF)^{1/3} = O(\eps^{1/3})$.

Together with the previous results, eqs.~\vepspreds, these
expressions imply that $\Sce_\SC^2(x)$ is quadratic in the
variable $\mu^2 \propto x^{2/3}$. Similarly, the gap in the SC
phase at optimal doping is related to the AF gap at zero doping
by:
\label\newvepsratio
\eq
\Sce_\SC(\opt) = {\mu_\opt \over \sqrt{2} \mu_\AF} \;
\varepsilon_\AF(0) = O(\eps^{1/3} J).
\eeq
This last equation, together with the interpretation of the
41 meV state as the pseudo-Goldstone boson of the SC phase,
and the underlying electronic scale $J \simeq 0.1 \, \meV$,
give the order of magnitude of the symmetry-breaking
parameter: $\eps \simeq 1\%$.
\endtopic

\section{Response Functions}

In order to measure the properties of these pseudo-Goldstone
particles, it is necessary to understand how they contribute
to the spin and electromagnetic response functions of the
materials. This is the topic of the present section.
Because these particles are weakly coupled, their response
may be computed perturbatively.

As we shall see, it shall become important for these purposes
also to understand how the pseudo-Goldstone states couple
to the other degrees of freedom in the low-energy system.
For this reason we also write down the electron/pseudo-Goldstone
particle couplings in this section.

The starting point for calculating the response functions is to
identify the dependence on the (pseudo-) Goldstone bosons of
the spin and electromagnetic currents. These are very easily
obtained, to lowest order in the derivative expansion,
by constructing the corresponding Noether currents using
the Lagrangian of eq.~\newinvform:
\label\Ncurrents
\eq
\matrix{
\rho_\em = - f_t^2 \Bigl( 1 + 2A \Bigr) \; \nq Q \;\dt\nq, \cr
\bfj_\em^\parallel = f_\parallel^2 \Bigl( 1+2D_\parallel
\Bigr) \; \nq Q \;\nabla \nq, \cr
j_\em^z = f_\perp^2 \Bigl( 1+2D_\perp \Bigr) \;
\nq Q \;\nabla_c \nq, \cr} \qquad
\matrix{
\vec\rho_\spin = f_t^2 \Bigl( 1 + 2B \Bigr) \; {\vec n}_\ssS \times
\;\dt {\vec n}_\ssS, \cr
\vec{\bfj}_\spin^\parallel = - f_\parallel^2 \Bigl( 1+2E_\parallel
\Bigr) \; {\vec n}_\ssS \times \;\nabla {\vec n}_\ssS, \cr
\vec{j}_\spin^z = - f_\perp^2 \Bigl( 1+2E_\perp \Bigr) \; {\vec n}_\ssS
\times \;\nabla_c {\vec n}_\ssS . \cr}
\eeq

Because the pseudo-Goldstone bosons are weakly coupled, correlations
of these currents may be be directly evaluated for free bosons, $\nq$
and $\ns$, plus perturbative corrections. This perturbative evaluation
conveniently organizes the contributions to the response
according to which states are responsible. The following
sections give some examples of such calculations.

\subsection{High-Energy Contributions}

Even though the effective lagrangian only contains as degrees of
freedom the states which actually appear in the low-energy spectrum,
it nonetheless carries the information as to how states at higher
energies contribute to response functions. The contributions of
higher-energy states are incorporated as they are integrated out
to produce the effective lagrangian itself, and so their contribution
to correlation functions may be read directly from the lagrangian.

More precisely, imagine coupling to a long-wavelength
electromagnetic field, $A_\mu$, or to a fictitious field, $s^a_\mu$,
coupling to spin, by making the substitution: $\partial_\alpha \nq
\to (\partial_\alpha - ieQ A_\alpha)\, \nq$ or $\partial_\alpha \ns
\to (\partial_\alpha - i T_a s^a_\alpha) \, \ns$, with  $T_a$ and
$Q$ the matrix generators of \three\ and \two, as
defined in eq.~\charges. The high-energy contribution to the
electromagnetic and spin response may then be obtained by
differentiating the effective Lagrangian twice with respect to
$A_\mu$ or $s^a_\mu$. The resulting correlation is proportional to
$\delta^d(\bfx - \bfx') \; \delta(t-t')$, as would be expected for a
mode which fluctuates on a scale much shorter than that over
which the response is computed.

Making this substitution for the electromagnetic response leads
to the magnetic penetration depth along the planes, $\lambda$,
and the electric screening length (or compressibility), $a$.
These are given by:
\label\screenlengths
\eq
\left( {1 \over \lambda^2} \right)_{\rm h.e.}
= 4 \pi \mu_0 e^2 q^2 f_\parallel^2 (1 + 2D) \; \nq \cdot \nq,
\qquad \left( {1\over a^2} \right)_{\rm h.e.}
= 4 \pi e^2 q^2 f_t^2  (1 + 2A) \; \nq \cdot \nq,
\eeq
where $\mu_0$ is the magnetic permeability of the
material.  The result in the ground state is obtained
by using the ground-state configurations: $\nq \cdot \nq = 1$
(SC) or $\nq \cdot \nq = 0$ (AF).

Similarly, the spin response obtained in this way gives the high-energy
contribution to the paramagnetic susceptibility:
\label\hesusc
\eq
\Bigl( {\chi}_{ab} \Bigr) _{\rm h.e.} = \mu_s^2 f_t^2 (1 + 2 B) \;
\ns T_a T_b \ns ,
\eeq
where $\mu_s$ is the magnetic moment of the pseudo-Goldstone
bosons.
These depend on temperature, only through the weak dependence
of coefficients $f_t$, $f_\parallel$, $A$, $B$ and $D$.

\subsection{Goldstone Poles}

\fig\GBpole

The next simplest contributions to response functions to compute are
the poles which occur in the correlation of the currents for
spontaneously broken (approximate) symmetries, due to
the contribution of the corresponding (pseudo-) Goldstone states.
This includes the superconducting contribution to the
electromagnetic response in the superconductor, and the magnon
contribution to the spin response in the antiferromagnet, in addition
to the pseudo-Goldstone boson couplings to the additional \five\
generators which rotate the spin and charge degrees of freedom
into one another. These are given in perturbation theory by the
Feynman graph of Fig.~\GBpole, which describes the direct
creation and destruction of the Goldstone boson from the
ground state by the current of interest.

\centerline{\phantom{This fill}
{\epsfxsize=7.5cm\epsfbox[-100 530 550 750]{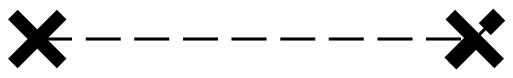}}}

\bigskip
\centerline{\bf Figure 3}
\medskip\noindent
\vbox{\baselineskip=10pt \eightrm The Feynman diagram which produces the
Goldstone pole contribution to the current-current correlation
function. The dashed lines represent pGB propagation.}
\bigskip

The result for the SC electromagnetic and the AF spin response,
obtained by summing the pole and the high-energy contributions,
is:
\label\GBpoleresult
\eq
\eqalign{
-i \theta(t) \Avg{[\rho_\em(x,t), \rho_\em(0)] }_{\sss GB} &=
\int {d^2 k \, d\omega \over (2\pi)^2 }\, e^{-i\omega t + i k\cdot x}
\left[ {4 \pi e^2 q^2 (1 + 2A) f^2_t  k^2 c_{sc}^2
\over   -\omega^2 + k^2 c_{sc}^2 - i \gamma_{sc} \omega}
\right] , \cr
-i \theta(t) \Avg{[\rho^a_\spin(x,t), \rho^b_\spin(0)] }_{\sss GB} &=
\int {d^2 k \, d\omega \over (2\pi)^2 }\,
e^{-i\omega t + i k\cdot x} \left[ {(1 + 2B) f^2_t  k^2 c_{af}^2
\;  (\ns T_a \eta) (\eta T_b \ns )
\over -\omega^2 + k^2 c_{af}^2 - i \gamma_{af} \omega} \right]
,\cr}
\eeq
where $c_{sc} \equiv c_\GB(SC)$ and $c_{af} \equiv c_\GB(AF)$.
$\gamma_{sc}$ and $\gamma_{af}$ represents
small damping contributions to the pGB dispersion relations.
For the spin-spin correlation, $\eta$ denotes the
normalized eigenvector of the gap matrix corresponding
to the GB direction whose pole is being computed. Notice
eqs.~\GBpoleresult\ reproduce eqs.~\screenlengths\
and \hesusc\ in the limit $\omega, k \to 0$.

\ref\SCGBpolexp{T. Timusk and D.B. Tanner,
in {\it Physical Properties of High-Temperature
Superconductors}, Volume $I$, edited by D.M. Ginsberg, World
Scientific, 1989.}

\ref\AFGBpolexp{J.M. Tranquada, \etal, \prb{40}{89}{4503};\bk
S. Shamoto \etal, \prb{48}{93}{13817}.}

Of course, eqs.~\GBpoleresult\ are not specific to \five-invariant
physics, since they also hold for  {\it any} antiferromagnet or
superconductor. This is because they simply contain the implications
of the Goldstone bosons associated with spontaneously
broken \three\ or \two\ invariance. As a result, although they provide
a good description of the response in these phases \SCGBpolexp,
\AFGBpolexp, \surveytexts\ this is not a real test of \five\
symmetry.\foot\alsotrue{For the same reasons, neither are such
responses detailed tests of the explicit models in which they are
usually derived.} It is the contributions of the
{\it pseudo}-Goldstone states in both phases which provide more
interesting information. After pausing to address a puzzle
concerning the pseudo-Goldstone pole in the spin response
of the SC phase, we close by describing some of the features
of the pGB response.

\subsection{Pseudo-Goldstone Poles: A Puzzle}

It was the contribution to neutron scattering of the spin-triplet
pGB state in the SC phase which originally motivated the \five\
picture. Since neutrons couple to electron spins, the \five\
interpretation of the neutron-scattering experiments requires
the pGB to contribute a resonance to the spin response functions.

This immediately leads to a puzzle. The spin-triplet pGB
contributes a pole similar to eqs.~\GBpoleresult\ to three \five\
currents (Zhang's $\pi$ operators) which are spontaneously
broken in the SC phase. But  Fig.~\GBpole\ does {\it not} produce
a pole in the SC spin response function in the SC phase,
because the symmetry of spin rotations is not broken in this phase.
The puzzle is how such a pseudo-Goldstone pole can arise
as a resonance in the SC spin correlation function.  This section
sketches how this puzzle is resolved within the effective field theory
framework.

\fig\SCpGBpole

The difficulty with producing a pGB pole in the spin correlation
function lies in the observation that $\vec\rho_\spin$ of
eqs.~\Ncurrents\ involves only {\it even} powers of the
boson field, $\ns$. The same is true of the interactions in the
effective lagrangian, eq.~\newinvform, and so it is difficult to
generate a graph of the form of Fig.~\SCpGBpole, which would
generate a resonant contribution to the spin correlation function.
The `blobs' of Fig.~\SCpGBpole\ represent any graphs which can
produce the triplet state starting from quasiparticles created
by the spin density, $\vec\rho_\spin$.

In our opinion the resolution of this puzzle comes from the
couplings of the pseudo-Goldstone bosons to the electronlike
quasiparticles in the SC phase. As stated earlier,
the existence of these quasiparticles
can be inferred from the evidence for \dwave\ pairing, since some of
these experiments indicate the existence of gapless excitations in
the SC phase. These excitations must be {\it in addition} to the four
Goldstone and pGB states of the SC phase.

\centerline{\epsfxsize=7.5cm\epsfbox[-100 530 550 750]{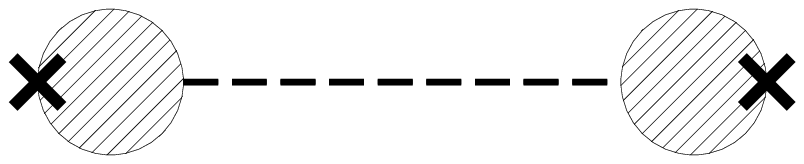}}

\bigskip
\centerline{\bf Figure 4}
\medskip\noindent
\vbox{\baselineskip=10pt \eightrm The Feynman diagram which produces the
pole contribution to the spin correlation function in the SC
phase due to the spin-triplet pseudo-Goldstone state. The `blobs'
represent fermion loops, while the dashed line represents
pseudo-Goldstone boson propagation.}
\bigskip

\ref\joeandco{J. Polchinski,
{\it Effective Field Theory of the Fermi Surface}, in {\sl
Recent Developments in Particle Theory, Proceedings of the 1992 TASI},
eds. J. Harvey and J. Polchinski (World Scientific, Singapore, 1993);\bk
R. Shankar, {\it Renormalization Group Approach
to Interacting Fermions}, \rmp{66}{94}{129}; \bk
T. Chen, J. Fr\"ohlich and M. Seifert,
{\it Renormalization Group Methods: Landau-Fermi Liquid
and BCS Superconductor}, preprint cond-mat/9508063.}

\ref\mynotes{C.P. Burgess, {\it op. cit}.}

To see how these quasiparticles can help with the puzzle,
suppose them to have the quantum numbers of electrons,
and to have couplings that are weak, so that their propagation
is approximately described by the Lagrangian density
$\Scl_0 = \int d^dp \; c^\dagger_p \bigl[ i \dt - E_p \Bigr] c_p$.
Here $c_p = \pmatrix{ c_{p\uparrow} \cr c_{p\downarrow} \cr}$
destroys a quasiparticle which propagates with dispersion
$\omega = E_p$, as is described by $\Scl_0$.
Standard arguments \joeandco, \mynotes\ can now be used to identify
those interactions which are the most important in the long-wavelength
limit. Goldstone boson couplings are all irrelevant (in the
renormalization group sense) in this limit, but the least
irrelevant of their couplings to the electronic quasiparticles
involve the emission and absorption of a single Goldstone particle.
The resulting electron Lagrangian density which describes this
is:
\label\elint
\eq
\Scl_{\rm int} = \int d^dp \, d^dk \; \Bigl[
g_\ssS(p,k) \; c^\dagger_{p+k} \vec\sigma c_p \; \cdot
\vec{n}_\ssq(k) + g_\ssq(p,k) \; c^\dagger_{p+k} (i \sigma_2)
c^*_{-p} \; \nq(k) \bigr] + \hc .
\eeq
This interaction
is least irrelevant for special regions of momentum of the
quasiparticle pairs \joeandco, such as when the net momentum
of the pair is close to zero.

Notice that an expectation value for $\nq$ introduces a quasiparticle
gap, proportional to $g_\ssq \Avg{\nq}$, so a \dwave\ symmetry
of the gap  restricts how $g_\ssq(p,k)$ can depend on momenta
lying on the Fermi surface.  Approximate
\five\ invariance relates the coupling $g_\ssq(p,k)$ to $g_\ssS(p,k)$,
and so implies a similar \dwave\ symmetry for $g_\ssS(p,k)$.

The principal observation at this juncture is that these
quasiparticles contribute quadratically to the spin density,
$\vec\rho_{\rm el spin} \propto \int d^dp \; c^\dagger_p
\vec\sigma  c_p$,
and so the coupling of eq.~\elint\ introduces a pGB pole
into the spin correlation function through the
Feynman graph of Fig.~\SCpGBpole, with the `blobs' representing
quasiparticle loops. Even though this graph does not arise
at leading order in perturbation theory, the singular shape of
the pole permits it to dominate the lower orders
for energies and momenta which
are related by the pGB dispersion relation.

\subsection{Pseudo-Goldstone Boson Response}

\fig\oneloop

Away from special features, such as the poles just
discussed, the dominant contribution made by pseudo-Goldstone
bosons to response functions arises through the Feynman graph
of Fig.~\oneloop.

\centerline{\phantom{T}
{\epsfxsize=7.5cm\epsfbox[-100 530 550 750]{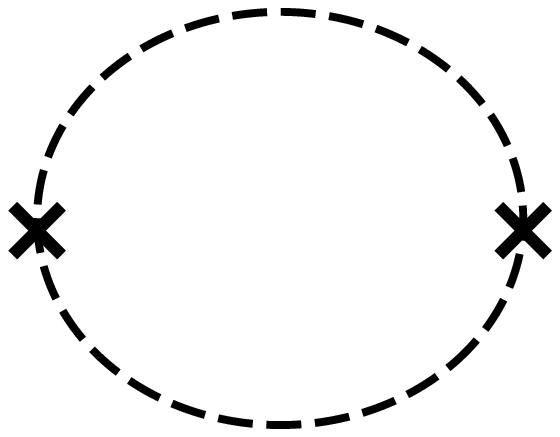}}}

\bigskip
\centerline{\bf Figure 5}
\medskip\noindent
\vbox{\baselineskip=10pt \eightrm The Feynman diagram which produces
the pseudo-Goldstone boson contribution to the
electromagnetic response in the AF phase.}
\bigskip

\ref\bosegas{J.I. Kapusta, \prd{41}{90}{6668}; \prd{46}{92}{4749}; \bk
V.L. Eletsky, J.I. Kapusta and R. Venugopalan, \prd{48}{93}{4398}.}

At zero frequency and momentum transfer this graph contributes
a temperature- and doping-dependent contribution to
the response functions, which are precisely those due to the
noninteracting gas of bosons having the dispersion relation
of the pseudo-Goldstone states. Since this dispersion relation
is `relativistic', the results are those of a gas of relativistic
bosons \bosegas.

In the SC phase one finds in this way the thermal paramagnetic
susceptibility due to the spin-triplet pGBs:
\label\SCsusc
\eq
\chi_{\rm pGB}(SC) = {2 \mu_s^2
\Omega_d \over (2 \pi c)^d} \int_0^\infty
dx \; {x^{d-3} \over E} \, \Bigl[x^2 +  (d-2) E^2 \Bigr] \;
n_\ssb(\mu=0) .
\eeq
Recall $\Omega_d$ is the solid angle swept out by a vector in
$d$ spatial dimensions ($\Omega_2 = 2\pi$, $\Omega_3 = 4\pi$),
and the boson dispersion relation is $E^2 = x^2  + \Sce^2$, for $x=kc$.
As in previous sections, $n_\ssb(\mu) = \Bigl[ e^{\beta(E-eq\mu)} - 1
\Bigr]^{-1}$ denotes the Bose-Einstein distribution function.
The thermal electric screening length due to the charged pGBs of
the AF phase is given by a very similar expression:
\label\AFscreen
\eq
\left( {1 \over a} \right)_{\rm pGB}(AF) =
{4 \pi e^2 q^2 \Omega_d \over (2 \pi c)^d} \int_0^\infty
dx \; {x^{d-3} \over E} \, \Bigl[x^2 +  (d-2) E^2 \Bigr] \;
\Bigl[ n_\ssb(\mu) + n_\ssb(-\mu) \Bigr] .
\eeq

Both $\chi_{\rm pGB}(SC)$ and $\left(1/a\right)_{\rm pGB}(AF)$
are therefore seen to be exponentially activated,
$\propto e^{-\beta \Sce}$, for $kT \ll \Sce$, and to vary as $T^{d-1}$
for $kT \gg \Sce$. Their contribution to the specific heat
per unit volume, $c_\ssv$, of the corresponding phases is
also exponentially small for $kT \ll \Sce$, and varies as $T^{d}$
for $kT \gg \Sce$. Unfortunately, the exponential suppression
makes this $T$ dependence difficult to detect at low temperatures,
while the large-$T$ power-law behaviour only applies, as derived,
for $T$ much greater than the pGB gap, and yet small enough that the
sample remains in the ordered phase.

\ref\boseemresp{C. Gale and J.I. Kapusta, \npb{357}{91}{65};\bk
A.I. Titov, T.I. Gulamov and B. K\"ampfer, \prd{53}{96}{3770}.}

The dynamic response function of a relativistic bose gas is also
known for nonzero frequencies and momenta \boseemresp.
This carries considerably more information about the
pseudo-Goldstone boson response, although the differentiation of
the pGB contributions from other degrees of freedom is easiest
at low frequencies. We close by presenting some preliminary
remarks concerning this response, and defer a more detailed
applications to experiments to a later publication.

Because of its electric charge, the pGB state of the AF
phase cannot contribute to a diagram of the form of
Fig.~\SCpGBpole, and so give a pole at low temperatures in the
electromagnetic response. Fig.~\oneloop\ nevertheless does
produce some strong dependence on frequency, due to
the singularity it implies near the threshhold for
producing pairs of pGBs. Below this threshhold the
electromagnetic response from Fig.~\oneloop\ is purely
real, while it is complex above. As a result, for temperatures
$T \gsim \Sce_\AF$ the pGBs contribute zero to
the conductivity for frequencies below threshold,
$\omega \lsim 2 \Sce_\AF$, but the conductivity then
grows steeply beyond this threshold. The implications of the resulting
expressions for electromagnetic scattering from cuprates in the
AF phase will be reported elsewhere.

\subsection{The Disordered Phase}

\ref\NMRstuff{A. Sokol and D. Pines, \prl{71}{93}{2813}.}

Many of the results obtained above for pseudo-Goldstone bosons
deep in the ordered phases might also be expected to apply
in the disordered phase. If so, the wealth of experiments available
there would permit many more detailed tests of \five\ symmetry.
Applications to the normal phase are also theoretically appealing,
since a number of striking features might be expected within
the \five\ picture, including two of the more striking
implications pointed out in ref.~\zhang: the explanation for
the pseudogap, and of the connection to a successful scaling
analysis of the temperature dependence of NMR relaxation times
\NMRstuff. Furthermore, the absence of spontaneous breaking
of \five\ also implies that the boson contribution
to correlation functions of the \five\ currents,
$j^\mu_a(x)$ -- with $a=1,\dots,10$ labelling the \five\
generators -- are very simply related in the \five-invariant
limit: $\Avg{j^\mu_a \;  j^\nu_b} \propto \delta_{ab}$.

In this section we make some {\it caveats} concerning the use
of the effective lagrangians of previous sections in the disordered
phase.
Our main point is to emphasize that conclusions drawn
from the effective lagrangian involving $\nq$, $\ns$ (and possibly
electronic quasiparticles) are {\it not} protected in the disordered
phase by the general low-energy constraints of Goldstone's theorem,
and so are necessarily more dependent on assumptions made
about the details of the underlying electronic interactions.
Although this makes
these predictions no longer simply consequences of the
symmetry-breaking pattern, they can nevertheless be worthwhile
as sources of more detailed information about this underlying
microscopic physics.

We next describe some of the ways in which model dependence
can enter predictions made for the disordered
phase using the boson lagrangians described in this paper.

\topic{1. Degrees of Freedom}
First, the
system's real degrees of freedom need not be as assumed,
since no general principles require the existence of low-energy
bosonic states described by $\nq$ and $\ns$ if no spontaneous
symmetry breaking occurs. An exception might be in the
immediate vicinity of the transition line into one of the
ordered phases, since continuity at this line would require
the gap for the pseudo-Goldstone states of the disordered phase
to still be small in the ordered phase. The deeper one moves
into the disordered phase, the less one would generically expect
the boson gaps to remain small compared to the intrinsic scale, $J$.

Furthermore, since the
four pseudo-Goldstone states should fill out a linear representation
of the unbroken \five\ of the disordered phase, fluctuations
in the modulus of the five-dimensional vector, $\bfn$, should
also appear in the low-energy spectrum. This argues that the
effective lagrangian of interest is of the form considered here
for  the free energy, (eq.~\invform\ supplemented by
time-derivative terms)  rather than as was used
for the pseudo-Goldstone bosons of the ordered phases
(eq.~\newinvform).
These differ through the relaxation of the constraint $\ns\cdot
\ns + \nq \cdot \nq = 1$, due to the fluctuations in the
magnitude of \five-breaking order parameter, $\bfn$.

\topic{2. Weak Coupling}
Even if the boson degrees of freedom exist at low energy,
they need not be weakly coupled since (unlike Goldstone bosons)
they are not required to decouple at low energies. It is noteworthy,
however, that for weakly-coupled systems the electron-boson
interactions given in eq.~\elint\ are the only couplings between
the electrons and bosons which can be marginal or relevant
(in the RG sense of refs.~\joeandco). By contrast,
there are a number of self-couplings among the bosons
which can be relevant or marginal in the infrared.
Of course, the existence of strong couplings
among the low-energy degrees of freedom doesn't necessarily
invalidate the use of the effective lagrangian, it could
just complicate the extraction of its predictions.

\topic{3. Bose-Einstein Condensation}
Even if the previous assumptions should apply to a particular
system, it is still true that a weakly-coupled version of
electrons and bosons cannot provide a good
description for the cuprates in the disordered phase
for dopings larger than
optimal. This is because
if the bosons $\nq$ are supposed to appear at low energies, and if
their couplings are weak, then the relation, $\mu(x)$, between
chemical potential and doping should be reasonably well described,
as in \S3.3, by a gas of free bosons and electrons. But this
description always implies Bose-Einstein condensation for
sufficiently large dopings, since for large enough $x$ the
bosons always `win' and, by condensing, dominate the expression for
the electric charge density.  This cannot describe the
observed {\it decrease} of the critical superconducting
temperature with increasing doping, above optimal doping.
\endtopic

Further work is necessary to better explore these implications for
the normal phase, and to more clearly identify which predictions
of the effective lagrangian for this phase are model-specific,
and which are more robust consequences of the symmetry-breaking
pattern.


\bigskip
\centerline{\bf Acknowledgments}
\bigskip

We would like to thank Charles Gale, Richard MacKenzie,
Rashmi Ray for helpful discussions, as well as J. Irving Kapusta
and Louis Taillefer for their guidance through the literature.
We thank Shou-Cheng Zhang for sharing his ideas with us about
high $T_c$ superconductors. This research was partially funded by the
N.S.E.R.C. of Canada, F.C.A.R du Qu\'ebec and the Norwegian Research Council.


{\baselineskip 15pt
\listrefs
}



\bye